\documentclass[11pt,superscriptaddress]{article}

\usepackage{tikz}
\usetikzlibrary{arrows.meta}
\usepackage{amsmath}
\usepackage{xcolor}
\definecolor{qpblue}{HTML}{2B3A67}
\definecolor{qpred}{HTML}{C44E52}
\definecolor{axisgray}{HTML}{444444}
\definecolor{regionA}{HTML}{E9C46A}
\usepackage[a4paper, total={5.8in, 9in}]{geometry}
\usepackage[indent,skip=4pt]{parskip}
\usepackage{amsmath,amssymb,amsfonts,amsthm,mathrsfs,mathtools,cases}
\usepackage{dsfont}
\usepackage{authblk}
\usepackage[colorlinks=true,linkcolor=blue, citecolor=blue, bookmarks]{hyperref}
\usepackage{cleveref}
\usepackage{microtype}
\usepackage{comment}
\usepackage{graphicx,adjustbox} 
\usepackage{subfig}
\usepackage{wrapfig}
\usepackage[font=small,labelfont=bf]{caption}
\usepackage[usenames,dvipsnames]{xcolor}  
\usepackage{booktabs}
\usepackage{braket}
\usepackage[nottoc]{tocbibind}
\usepackage{nicematrix}
\usepackage{tikz}
\usepackage{pgfplots}
\usepackage{cite}
\pgfplotsset{compat=1.18}
\usepackage{enumitem}

\usetikzlibrary{matrix}
\usetikzlibrary{decorations.markings}
\usetikzlibrary{patterns}
\usetikzlibrary{arrows.meta}
\usepackage{color,ifthen,amsthm,amsmath,amsxtra,amsfonts,dsfont,graphicx,bm,tikz,scalerel,wasysym,bbm,graphicx,amsthm,braket,physics,empheq,CircuitTikz,float}

\numberwithin{equation}{section}

\crefname{figure}{Fig.}{Figs.}
\crefname{equation}{Eq.}{Eqs.}
\crefname{section}{Sec.}{Secs.}
\crefname{appendix}{Appendix}{Appendices}
  
\colorlet{darkerblue}{blue!80!black}
\colorlet{lightblue}{blue!70!white}
\usepackage{braket}
\hypersetup{
    colorlinks,
    linkcolor={darkerblue},
    citecolor={darkerblue},
    urlcolor={darkerblue},
    backref=true
    }

\newcommand{\R}{{\mathbb{R}}}

\usepackage{CircuitTikz,braket}
\newcommand{\be}{\begin{equation}}
\newcommand{\ee}{\end{equation}}
\newcommand{\bea}{\begin{eqnarray}}
\newcommand{\eea}{\end{eqnarray}}
\newcommand{\bse}{\begin{subequations}}
\newcommand{\ese}{\end{subequations}}

\usepackage{amsmath}
\newcommand{\sign}{\text{sign}}
\usepackage{tikz}
\usepackage{graphicx}
\usepackage{xcolor}

\title{\bf From nonlocal correlations to local symmetry breaking: Asymmetry dynamics of crosscap states}
\author{Konstantinos Chalas$^{1,2}$, Pasquale Calabrese$^{1}$ and  Colin Rylands$^3$}

\date{}

\begin{document}
\maketitle
{\small
\vspace{-5mm}  \ \\
{$^{1}$}  SISSA and INFN, via Bonomea 265, 34136 Trieste, Italy\\[0.1cm]
\medskip
{$^{2}$} Dipartimento di Fisica e Astronomia, Università di Bologna and INFN, Bologna, Italy\\ 
{$^{3}$} Centre for Fluid and Complex Systems, Coventry University, Coventry, CV1 2TT, 
UK
%\\[-0.2cm]
%\medskip \hspace{0.25 cm }United Kingdom\\[-0.1cm]
}

\maketitle
\begin{abstract}

Under unitary dynamics preserving a global symmetry, bipartite entanglement typically grows while an initially broken symmetry is locally restored. 
Here, we show that this conventional scenario can be reversed when symmetry-breaking coherence is initially encoded in long-range entanglement.
We investigate this phenomenon in a family of  charge deformed crosscap states, characterized by antipodal entangled pairs with tunable entanglement and symmetry breaking. 
We first characterize their static properties, finding a Page-like entanglement profile and an entanglement asymmetry that vanishes for subsystems smaller than half the system size, but grows logarithmically for larger subsystems. 
For symmetric subsystems, we also analyze the symmetry-resolved entanglement and investigate how it evolves as the subsystem is continuously deformed away from a symmetric partition. Following a quench to a free-fermion tight-binding chain, we uncover an unconventional dynamics in which the entanglement entropy decreases whereas the entanglement asymmetry increases, so that initially symmetric subsystems dynamically develop local symmetry breaking. This behavior is quantitatively captured by a quasiparticle picture describing the propagation of antipodal correlations. Our results reveal how symmetry-preserving dynamics can redistribute globally encoded symmetry-breaking coherence, making it locally accessible, and establish the initial entanglement structure as a key ingredient in determining the fate of local symmetries out of equilibrium.

\end{abstract}
\tableofcontents

\section{Introduction}
\label{sec:crossasym:intro}
Bipartite entanglement entropy is one of the most widely used tools to investigate and characterize pure states of many-body quantum systems. In equilibrium,  it exhibits universal scaling  with subsystem size for ground states of local Hamiltonians~\cite{latorre_critical,plenio_virmani_ent_2005,amico2008entanglement,horodecki2009quantum,Calabrese_2009_1,arealaws,Calabrese_2009,Laflorencie_2016}. Out of equilibrium, it also displays universal behaviour: typically showing linear in time increase and eventual saturation, indicating local relaxation in quenches from lowly entangled states~\cite{calabrese2005evolution,alba2017entanglement,calabrese2020entanglement}. Discovering these universal behaviours and how they might be avoided has led to a far greater understanding of quantum matter. Going beyond this and exploring the interplay between entanglement and other system properties can then give a more refined picture of the structure of many-body quantum states. A natural candidate for this is to consider the relationship between the symmetries of a system and its entanglement and  ask how the entanglement of a state is reshaped when it respects, or breaks, a certain symmetry or how does the presence of entanglement affect symmetry restoration and breaking. 

A state $\rho$ is symmetric with respect to a $U(1)$ symmetry generated by a local charge $Q$  if $[Q,\rho]=0$. The reduced density matrix of a subsystem $A$, $\rho_A=\tr_{\bar A}[\rho]$ is then also symmetric if the charge can be split into disjoint pieces acting separately on $A$ and $\bar A$. That is, if $Q=Q_A+Q_{\bar A}$ with $Q_{\bar A}$ being the charge operator restricted to act only in ${\bar A}$, then $[\rho,Q]=0\Rightarrow [\rho_A,Q_A]=0$. Such a symmetric state is therefore block diagonal in the basis of the subsystem charge $Q_A$ and as a result we can decompose the entanglement entropy,  $S[\rho_A]=-\tr [\rho_A\log \rho_A ]$, into number and configuration entropy~\cite{SREE15Laflorencie_2014,SREE07GoldsteinSela,SREE06SierraEquipartition,SREE12Bonsignori_2019,SREE04Murciano_dissipation_2023,SREE05_mbl_entan },
\begin{eqnarray}
    \label{eq:crossasym:symm_resolution}S[\rho_A]=S_{\rm num}[\rho_A]+S_{\rm conf }[\rho_A]~.
\end{eqnarray}
The number entropy is defined as 
\begin{eqnarray}
    S_{\rm num}[\rho_A]=-\sum_qp(q)\log p(q),\quad\quad p(q)=\tr[\Pi_q\rho_A]
\end{eqnarray}
where $\Pi_q$ is the projector onto the eigenspace of $Q_A$ with eigenvalue $q$. It is the contribution to the entanglement coming from fluctuations of the charge in the subsystem. The configuration entropy is 
\begin{eqnarray}
    S_{\rm conf}[\rho_A]=\sum_qp(q)S[\rho_{A,q}],\quad\quad\rho_{A,q}=\frac{\Pi_q\rho_A\Pi_q}{p(q)}~.
\end{eqnarray}
Here $S[\rho_{A,q}]$ is the symmetry resolved entanglement and gives the entanglement of a given charge sector. The configuration entropy is thus the average entanglement of all the charge  sectors over the distribution $p(q)$. These quantities have been extensively studied both in and out of equilibrium with numerous intriguing properties uncovered~\cite{SREE15Laflorencie_2014,SREE07GoldsteinSela,SREE06SierraEquipartition,SREE14_qpp_sree_cft_ffs,SREE01Parez_2021,SREE02Piroli_2022,SREE03_fcs_sree_st_duality,SREE04Murciano_dissipation_2023,SREE05_mbl_entan,SREE12Bonsignori_2019,SREE18_vitale_purification,SREE19_ent_barrier,SREE_13_Castro_Alvaredo_2024}. For instance, in quenches from strongly symmetric states the system experiences a time delay whereby some charge sectors are inaccessible at short times as the requisite charge fluctuations have not had time to build up~\cite{SREE14_qpp_sree_cft_ffs}. Additionally, it has been observed, in equilibrium and after local relaxation following a quench, that entanglement is equipartitioned amongst the charge sectors i.e. $S[\rho_{A,q}]\simeq S[\rho_{A,q_0}]$ where $q_0$ is the average charge\cite{SREE07GoldsteinSela,SREE06SierraEquipartition,SREE12Bonsignori_2019,SREE14_qpp_sree_cft_ffs,capizzi_ruggier_excited_states_sree_cft_2020}.  

When a state is not symmetric this decomposition receives an additional contribution~\cite{travaglino2025quench}, 
\begin{eqnarray}
    \label{eq:crossasym:asymm_resolution}S[\rho_A]=S_{\rm num}[\rho_A]+S_{\rm conf }[\rho_A]-\Delta S[\rho_A]~.
\end{eqnarray}
This new term, alternately known as the entanglement asymmetry, relative entropy of asymmetry or just asymmetry for short~\cite{ares2023entanglement,vaccaro2008tradeoff,gour2009measuring}, is defined as 
\begin{eqnarray}
    \Delta S[\rho_A]=\tr[\rho_A\,(\log \rho_A-\log\bar\rho_{A,Q})],~\quad\quad\bar{\rho}_{A,Q}=\sum_q p(q)\rho_{A,q}~.
\end{eqnarray}
In the above formula the entanglement asymmetry can be interpreted as the reduction in entanglement caused by the presence of coherences between different charge sectors compared to a state with the same $p(q)$ and $\rho_{A,q}$. The entanglement asymmetry is more commonly studied in its own right as a faithful probe of symmetry breaking in many-body quantum systems~\cite{ares2023entanglement}. If it is non-zero, the state breaks the symmetry generated by $Q$ whereas if it is zero the state is symmetric. Moreover, it is an asymmetry monotone meaning that it is non increasing under symmetric quantum channels~\cite{gour2009measuring}. Thus its value gives a meaningful quantifier of symmetry breaking, larger $\Delta S[\rho_A]$ corresponds to more symmetry breaking. 
Physically, the asymmetry measures the distinguishability of  $\rho_A$ from the space of symmetric states. The state $\bar{\rho}_{A,Q}$ being the unique symmetric state which is least distinguishable from $\rho$~\cite{gour2009measuring}. The entanglement asymmetry has been widely studied in the last few years in a wide range of systems and settings~\cite{ares2023entanglement,ampc-23,ASYMM02bertini2023dynamics,fossati_2025_noninvertible_asymm,fossati2024entanglement,Khor_2024,ASYMM03Ferro_2024,gibbins2025,fadc-24,ASYMM05capizzi2023universal,castroalvaredo2026entanglement,mazzoni2025breaking,gotta2026enhancing,ares2026lower,florio2025entanglementasymmetrygaugetheories,Russotto2025b,Kusuki2024entanglement,vitouladitis_2025_asymmetry,Russotto_2026,hammer2026asymmetrydynamicsnonequilibriumsymmetrybreaking,vescovo2026,benedito2026,yamashika2026}, also experimentally~\cite{ASYMM07joshi2024observing,Joshi2026,xu2026}. 
The quantity is particularly insightful when studying the time evolution of a symmetry breaking state under symmetric dynamics. Within that setting, it is commonly seen that the subsystem asymmetry monotonically decreases in time, eventually  vanishing indicating local restoration of the broken symmetry~\cite{ares2023entanglement,rylands2024microscopic,ASYMM01murciano2023xympemba,Chalas_2024,das2026,fossati2026boundary,DiGiulio2025easurement,ASYMM07joshi2024observing,liu_mpemba_circuit-24,liu2024symmetry2,yac-24,deluca_qme_circuits2025,cma-24,summer2025resource,yamashika2024quenching,ares2025ead,kusuki2026,muller2026,Yamashika_ares_lr_qme_2026,Tamizhselvan2026,Calabrese_2026,aditya2026higher,McRoberts_2026}. Notably, although $\Delta S[\rho_A]$ is an asymmetry monotone, the reduced state does not necessarily evolve via a symmetric quantum channel even if the full state does. Therefore, there is nothing to prevent the subsystem asymmetry from actually increasing under global symmetric dynamics. 

In this paper we study the interplay of entanglement, symmetry and asymmetry in a single parameter class of long range entangled states called \textit{charge deformed \textit{crosscap} states}. Crosscap~\cite{ishibashi1988crosscap,caetano2022crosscap}, or entangled antipodal pair states as they are also called~\cite{yoneta_eap_PhysRevLett.133.170404,yoneta_eap_PhysRevResearch.6.L042062},  were first studied in quantum field theory, where they related to boundary conditions on non-orientable manifolds~\cite{ishibashi1988crosscap,fioravanti199sewing,tang2017universal,tang2018klein}.
Lattice, spin versions were then introduced, taking the form of translationally invariant states with correlations only between antipodal points on the chain~\cite{caetano2022crosscap,he2023integrable,gombor2022integrable1,gombor2021exact,gombor2022integrables2}, which can be realised as zero energy eigenstates of some local Hamiltonians~\cite{mestyan2025crosscapstatestunableentanglement}, while similar states have been also discussed in the same context in~\cite{hokkyo2026exactthermalstabilizereigenstates}. Scar states with similar structure have also been found~\cite{ivanov_2025_scar_PRL,mougdalya_volume_law_2025}.
Moreover their fermionic versions were discussed in~\cite{zhang2024crosscap,chalas2024quenchdynamicsentanglementcrosscap,chalas2025entanglementevolutionentangledmultipodal}, where in~\cite{chalas2024quenchdynamicsentanglementcrosscap,chalas2025entanglementevolutionentangledmultipodal} the dynamics of entanglement entropy after a free fermionic quench was studied.
Quenches from crosscap states in field theories have also been studied~\cite{wei2025crosscapquenchesentanglementevolution,bai2025spatially,chen2025exact}.
These states have garnered attention due to their novel  properties, in particular, they are locally indistinguishable from the maximally mixed state meaning that their entanglement entropy follows a volume law~\cite{caetano2022crosscap}, which is similar to the Page curve of Haar random states~\cite{page1993average}, with the distinction that in this case the result is not only valid in the thermodynamic limit and it is true at the level of one pure state and not for an ensemble.  
They are among the few highly entangled initial states whose nonequilibrium dynamics has been investigated analytically (see Refs.~\cite{bucciantini2014quantum,kormos2014stationary,sierra_exotic_correlation_spread_2022,santalla2023entanglement,sabadini2026} for other examples).

We compute the entanglement, its symmetry resolution and the entanglement asymmetry of the charge deformed crosscap states both statically and dynamically out of equilibrium. We show that the degree of the charge deformation tunes both entanglement and the symmetry breaking of the states.  Moreover, the states are locally symmetric for subsystem sizes which are less than half the system size, irrespective of their charge deformation. For larger subsystems, the state breaks the symmetry, displaying some similarities to Haar-random states~\cite{ampc-23,Russotto_2025}. However, it notably does not exhibit the sharp jump in entanglement asymmetry observed in Haar-random states~\cite{ampc-23,Russotto_2025}.
We then study the dynamics of these states when quenched with the tight binding chain and present a quasiparticle picture for the entanglement and asymmetry dynamics for subsystems that are both smaller and larger than half system size. Using this we find that upon evolving for certain subsystem sizes, an initially symmetric subsystem becomes asymmetric. Thus, in contrast to previously seen behaviour wherein symmetric dynamics leads to symmetry restoration, we find the opposite, namely symmetry breaking via symmetric dynamics. 

The rest of the paper is organised as follows. In Section~\ref{sec:crossasym:crosscap} we introduce the fermionic crosscap states, their charge deformation, and compute the reduced density matrix for arbitrary subsystem size. In Section~\ref{sec:crossasym:equilibrium} we discuss the static properties of our states, computing the entanglement entropy and the asymmetry. We also examine the relation between these quantities as a function of subsystem size. In Section~\ref{sec:crossasym:quecnh} we introduce our quench protocol and recap the quasiparticle picture for entanglement. In Section~\ref{sec:crossasym:asymm} we extend this analysis to the entanglement asymmetry and compare our results to exact numerics. In the last section~\ref{sec:crossasym:conclusions} we conclude.

\section{Charge deformed crosscap states}\label{sec:crossasym:crosscap}

In this section, we will introduce a deformed version of the fermionic crosscap states considered in \cite{chalas2024quenchdynamicsentanglementcrosscap}, which feature a tunable amount of entanglement in the state.

\subsection{System and states}
We consider a system of $\mathcal{L}=2L$ sites with $c^\dagger_x, c_x$ being fermionic creation and annihilation operators for site $x=1, \dots ,2L$, and the vacuum state $\ket{0}$ which obeys $c_x\ket{0}=0$ for all $x$. These operators obey canonical anti-commutation relations $\{ c^\dag_x,c_y\}=\delta_{x,y},~\{c_x,c_y\}=0$ and in addition we impose anti-periodic boundary conditions $c_{x+2L}=-c_x$. Within this system we will be interested in studying the properties of a subsystem $A$ which we take to be formed of $|A|=\ell$ contiguous sites with $1\leq \ell \leq 2L$.

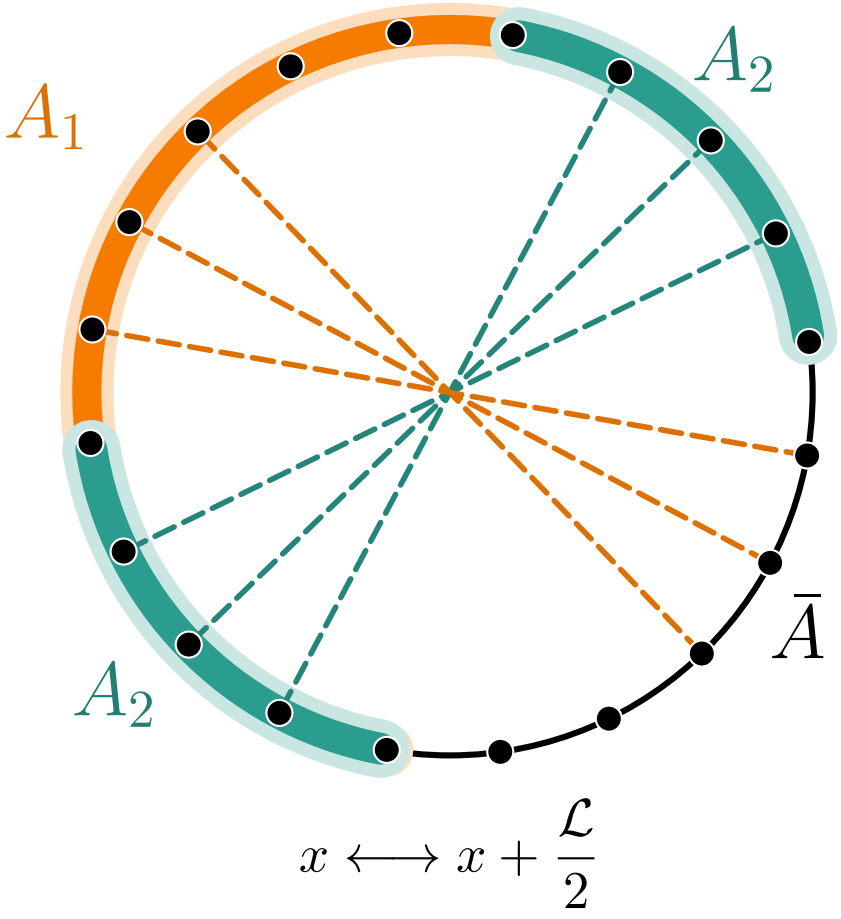
\begin{figure}[t]
\centering

\begin{minipage}[c]{0.95\linewidth}
\centering

\definecolor{suborange}{HTML}{F57C00}
\definecolor{specialteal}{HTML}{2A9D8F}
\definecolor{guidegray}{HTML}{666666}

\begin{tikzpicture}[
    line cap=round,
    line join=round,
    scale=1.0,
    every node/.style={transform shape}
]

% ====================================================
% Parameters
% ====================================================

\def\R{2.40}

% Full subsystem A
\def\Astart{10}
\def\Aend{260}

% Centres of the two antipodal A_2 regions
\def\QPa{44}
\pgfmathsetmacro{\QPb}{\QPa+180}

% Angular half-width of A_2
\def\dA{35}

% Number of lattice sites
\def\N{20}

% ====================================================
% Main circle
% ====================================================

\draw[
    black,
    line width=1.15pt
]
(0,0) circle (\R);

% ====================================================
% Subsystem A
% ====================================================

% Orange glow
\draw[
    suborange!25,
    line width=10pt
]
(\Astart:\R)
arc[
    start angle=\Astart,
    end angle=\Aend,
    radius=\R
];

% Main orange subsystem arc
\draw[
    suborange,
    line width=5.5pt
]
(\Astart:\R)
arc[
    start angle=\Astart,
    end angle=\Aend,
    radius=\R
];

% ====================================================
% A_2 regions
% ====================================================

% Upper A_2 glow
\draw[
    specialteal!25,
    line width=11pt
]
({\QPa-\dA}:\R)
arc[
    start angle={\QPa-\dA},
    end angle={\QPa+\dA},
    radius=\R
];

% Upper A_2
\draw[
    specialteal,
    line width=6.0pt
]
({\QPa-\dA}:\R)
arc[
    start angle={\QPa-\dA},
    end angle={\QPa+\dA},
    radius=\R
];

% Lower A_2 glow
\draw[
    specialteal!25,
    line width=11pt
]
({\QPb-\dA}:\R)
arc[
    start angle={\QPb-\dA},
    end angle={\QPb+\dA},
    radius=\R
];

% Lower A_2
\draw[
    specialteal,
    line width=6.0pt
]
({\QPb-\dA}:\R)
arc[
    start angle={\QPb-\dA},
    end angle={\QPb+\dA},
    radius=\R
];

% ====================================================
% Lattice sites
%
% theta_j = 44 - 360*j/N
%
% With N=20, sites are separated by 18 degrees.
% Sites j and j+10 are antipodal.
% ====================================================

\foreach \j in {0,...,19}{
    \pgfmathsetmacro{\ang}{44-360*\j/\N}
    \coordinate (site\j) at (\ang:\R);
}

% ====================================================
% COMPLETE ANTIPODAL PAIRS -- A_2
%
% site 19 <--> site 9
% site  0 <--> site 10
% site  1 <--> site 11
%
% Both endpoints lie in A.
% ====================================================

\foreach \j/\k in {19/9,0/10,1/11}{
    \draw[
        specialteal!85!black,
        line width=1.05pt,
        dash pattern=on 4pt off 2pt
    ]
    (site\j) -- (site\k);
}

% ====================================================
% SPLIT ANTIPODAL PAIRS -- A_1 <--> \bar A
%
% site 3 <--> site 13
% site 4 <--> site 14
% site 5 <--> site 15
%
% One endpoint lies in A, the other in its complement.
% ====================================================

\foreach \j/\k in {3/13,4/14,5/15}{
    \draw[
        suborange!90!black,
        line width=1.05pt,
        dash pattern=on 4pt off 2pt
    ]
    (site\j) -- (site\k);
}

% ====================================================
% Lattice dots
% ====================================================

\foreach \j in {0,...,19}{
    \filldraw[
        fill=black,
        draw=white,
        line width=0.4pt
    ]
    (site\j) circle (2.45pt);
}

% ====================================================
% Labels
% ====================================================

% A_1
\node[
    font=\Large,
    anchor=east,
    text=suborange!90!black
]
at (-2.25,1.82)
{$A_1$};

% Upper A_2
\node[
    font=\Large,
    anchor=east,
    text=specialteal!80!black
]
at (2.30,2.20)
{$A_2$};

% Lower A_2
\node[
    font=\Large,
    anchor=east,
    text=specialteal!80!black
]
at (-1.80,-2.00)
{$A_2$};

% Complement
\node[
    font=\Large
]
at (2.30,-1.55)
{$\bar A$};

% ====================================================
% Physical interpretation
% ====================================================

\node[
    anchor=west,
    align=left,
    font=\small
]
at (3.25,0.65)
{
    \textcolor{suborange}{
        $\boldsymbol{A_1}$: split antipodal pairs
    }
    \\[-1pt]
    \hspace{0.30cm}
    one partner in $A$, one in $\bar A$
    \\[9pt]
    \textcolor{specialteal}{
        $\boldsymbol{A_2}$: complete antipodal pairs
    }
    \\[-1pt]
    \hspace{0.30cm}
    both partners in $A$
    \\[11pt]
    $A=A_1\cup A_2$
};

% ====================================================
% Antipodal relation
% ====================================================

\node[
    font=\small
]
at (0,-3.05)
{$x\longleftrightarrow
  x+\dfrac{\mathcal L}{2}$};

\end{tikzpicture}

\end{minipage}

\caption{
Initial-state sketch for a subsystem of size
$\ell>\mathcal L/2$.
Each lattice site $x$ is entangled with its antipodal partner
$x+\mathcal L/2$.
The subsystem decomposes as $A=A_1\cup A_2$.
The orange region $A_1$ contains split antipodal pairs, for which
one member belongs to $A$ while its partner belongs to $\bar A$.
The two teal regions forming $A_2$ are antipodal images of one
another and contain complete antipodal pairs, for which both members
belong to $A$.
Orange and teal dashed lines illustrate representative split and
complete antipodal pairs, respectively.
}
\label{fig:crossasym:schematic_antipodal}

\end{figure}

The fermionic crosscap state is  defined to be~\cite{chalas2024quenchdynamicsentanglementcrosscap}
\begin{eqnarray}
    \ket{\mathcal{C}}=\frac{1}{2^{L/2}}e^{\frac{1}{2}\sum_{x,y=1}^{2L}G_{x,y}c^{\dag}_xc^\dag_y}\ket{0}
\end{eqnarray}
where the $2L\times 2L$, antisymmetric matrix $G_{x,y}$ is given by 
\begin{eqnarray}
    G_{x,y}=\delta_{x+L,y}-\delta_{x,y+L},
\end{eqnarray}
and $\ket{0}$ is the vacuum containing no fermions. This state can be rewritten in a number of different ways,
\begin{eqnarray}
    \ket{\mathcal{C}}&=&\frac{1}{2^{L/2}}e^{\sum_{x=1}^Lc^{\dag}_{x}c^\dag_{x+L}}\ket{0}
 =\prod_{x=1}^{L}\Big[\frac{1+c^\dag_{x} c^\dag_{x+L}}{\sqrt{2}}\Big]\ket{0}~,
\end{eqnarray}
from which its properties can be extracted.
In particular, the fermionic crosscap state is a Gaussian, squeezed state featuring long range correlations between antipodal points of the system which has the geometry of a ring, see Figure \ref{fig:crossasym:schematic_antipodal}. In this paper, we will study a deformation of these states generated by the charge operator, 
\begin{eqnarray}
Q=\sum_{x=1}^{2L}c_{x}^\dag c_x~,
\end{eqnarray}
which simply counts the number of fermions in the system. With this, we define the charge deformed crosscap states as
\begin{eqnarray}
    \ket{\mathcal{C}_ \mu} &=&\frac{e^{\frac{1}{4} \mu Q}\ket{\mathcal{C}}}{||e^{\frac{1}{4} \mu Q}\ket{\mathcal{C}}||}
    = \frac{1}{(1+e^\mu)^{L/2}}e^{\frac{1}{2}\sum_{x,y}G^\beta_{x,y}c^{\dag}_xc^\dag_y}\ket{0}=\prod_{x=1}^{L}\Big[\frac{1+{\beta} c^\dagger_{x} c^\dagger_{x+L}}{\sqrt{1+{\beta^2}}}\Big]\ket{0}
\end{eqnarray}
where, for later convenience, we also introduce $\beta=e^{\mu/2},$ in terms of which the charge deformed crosscap states are 
particularly easy and the matrix that appears in the real-space expression is
\begin{eqnarray}\label{eq:crossasym:G_matrix}
    G^\beta_{x,y}=\beta(\delta_{x+L,y}-\delta_{x,y+L})~.
\end{eqnarray}
The deformation introduces a bias of the charge within the state such that for $ \mu\to-\infty$ it reduces to the vacuum state $\lim_{ \mu\to-\infty}\ket{\mathcal{C}_ \mu}= \ket{0}$ and in the opposite limit, $ \mu\to\infty $, it becomes the fully filled state with a single fermion on each site. At $ \mu=0$ we recover the standard crosscap state, $
    \ket{\mathcal{C}_0}=\ket{\mathcal{C}}.$ A similar deformation is used in the study of pure thermal states\cite{yoneta_eap_PhysRevResearch.6.L042062,yoneta_eap_PhysRevLett.133.170404} but where a local Hamiltonian is used as the deformation operator~\cite{chen2025exact}.  In  thermal pure states \cite{yoneta_eap_PhysRevResearch.6.L042062} the goal is to produce a state which exhibits the properties of a thermal state at the subsystem level but which remains pure. We will see below that the charge deformed crosscap states have similar features.  Similar states with antipodal correlations and non-maximal entanglement albeit in spin form, have also been studied in~\cite{mestyan2025crosscapstatestunableentanglement} and they have been shown to be exact mid-spectrum eigenstates of local spin chains\cite{ivanov_2025_scar_PRL,yoneta_eap_PhysRevLett.133.170404}.

\subsection{Reduced density matrix}
Since the deformed crosscap states are Gaussian, their properties can be completely understood through the two-point functions of $c_x, c^\dag_y$. To this end we introduce the $4L\times 4L$ covariance matrix whose elements are given by
\begin{eqnarray}
    \Gamma_{x,y}=2\begin{pmatrix}
        \bra{\mathcal{C}_ \mu} c^\dag_x c_y\ket{\mathcal{C}_ \mu} -\frac{1}{2}\delta_{x,y} & \bra{\mathcal{C}_ \mu} c^\dag_x c^\dag_y\  \ket{\mathcal{C}_ \mu}\\
        \bra{\mathcal{C}_ \mu} c_xc_y\ket{\mathcal{C}_ \mu} & \bra{\mathcal{C}_ \mu} c_xc^\dag_y  \ket{\mathcal{C}_ \mu}-\frac{1}{2}\delta_{x,y} 
    \end{pmatrix}~.
\end{eqnarray} Using the explicit form of the state $\ket{\mathcal{C}_ \mu}$ one can verify that the $\Gamma$  takes the form
\begin{eqnarray}
    \Gamma_{x,y}=2\begin{pmatrix}
        \left(\theta_ \mu-\small\frac{1}{2}\right)\delta_{x,y}& \sqrt{\theta_ \mu(1-\theta_ \mu)}\delta_{|x-y|,L}\sign(y-x)\\
        \sqrt{\theta_ \mu(1-\theta_ \mu)}\delta_{|x-y|,L}\sign(x-y) & \left(\small{\frac{1}{2}}-\theta_ \mu\right)\delta_{x,y}
    \end{pmatrix},
\end{eqnarray}
where $\theta_ \mu={e^{\mu}}/(1+{e^{\mu}})$ is the charge density in the state, i.e. $\theta_ \mu=\bra{\mathcal{C}_ \mu} Q\ket{\mathcal{C}_ \mu}/2L$. The eigenvalues of $\Gamma$ are thus highly degenerate and given by $\pm 1$, as expected for a pure state.

The reduced density matrix describing the subsystem is also Gaussian and so its properties are similarly governed by the same covariance matrix, albeit with the spatial indices restricted to lie inside $A$. We denote the restricted covariance matrix by
\begin{eqnarray}[\Gamma_A]_{x,y}=\Gamma_{x,y}~~~~~;x,y\in A~.
\end{eqnarray}
Generally, we will denote subsystem quantities with a subscript $A$, e.g. the subsystem charge is $Q_A=\sum_{x\in A}c^\dag_xc_x$~.  
The restricted covariance matrix is  completely diagonal for $\ell\leq L$ with an increasing number of off-diagonal elements for $\ell>L$. The simple form allows one to determine the eigenvalues explicitly. We find eigenvalues $\pm (2\theta_ \mu-1)$ each with degeneracies of min$(\ell,2L-\ell)$ and eigenvalues $\pm 1$ with degeneracies max$(0,2\ell-2L)$. This structure of eigenvalues hints that the reduced state factorises into a product of pure and mixed terms, with the former corresponding to the unit norm eigenvalues. This is similar to what is expected for the crosscap state written in terms of spins where the partial trace is easily taken.

For the fermionic system, taking the partial trace is more subtle due to the anti-commutation relations. Nevertheless, we can compute the reduced density matrix following the procedure of \cite{peschel_chung_2001} using fermionic coherent state path integrals. The result is
\begin{eqnarray}
   \rho_A= \frac{1}{(1+e^\mu)^{|A_2|/2}} e^{\frac{1}{2}\sum_{x,y\in A_2}G^\beta_{x,y}c^
    \dag_x c^\dag_y} \prod_{x\in A_2}(1-c^\dag_xc_x)e^{-\frac{1}{2}\sum_{x,y\in A_2}G^\beta_{x,y}c_x c_y} 
     \frac{e^{ \mu Q_{A_1}}}{\tr [e^{ \mu Q_{A_1}}]}
     \label{eq:crossasym:reduced_density}
\end{eqnarray}
where $G^\beta_{x,y}$ is given in~\eqref{eq:crossasym:G_matrix} and we have split the subsystem $A=A_1\cup A_2$ so that $A_1$ contains sites whose antipodal point is not in $A$ while $A_2$ contains the pairs of antipodal points which are contained in $A$, see Figure~\ref{fig:crossasym:schematic_antipodal}. The size of these regions coincides with the eigenvalue degeneracies of $\Gamma_A$, specifically $|A_1|={\rm min}(\ell,2L-\ell)$ and $|A_2|={\rm max}(0,2\ell-2L)$.  Here we see the approximate factorisation mentioned above. The rightmost term is a Gibbs state on the region $A_1$ which is mixed, whereas the leftmost and central terms take the form of the initial state itself, i.e. a fermionic crosscap state, but which exists only in $A_2$. One can confirm that this state is Gaussian and has the correct covariance matrix $\Gamma_A$ and therefore is the correct reduced density matrix. We use this form in the next section to calculate the properties of $\ket{\mathcal{C}_ \mu}$.

\section{Initial-state properties}\label{sec:crossasym:equilibrium}
Having introduced the charge deformed crosscap states in the previous section we now study their static entanglement entropy and entanglement asymmetry, starting with the former.

\subsection{Entanglement entropy}
\begin{figure}
    \centering
    \includegraphics[width=0.95\linewidth]{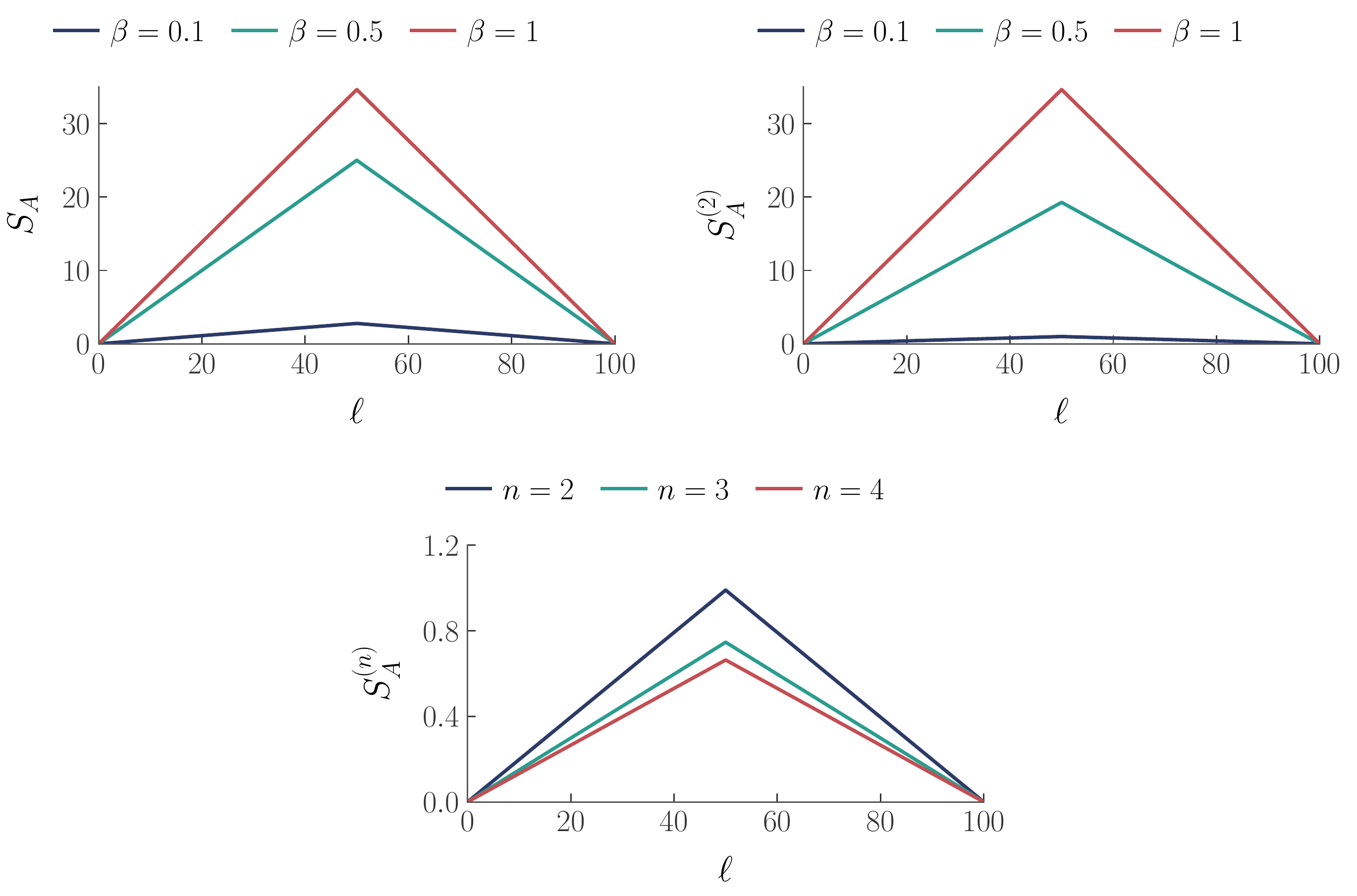}
    \caption{Entanglement entropy of the charge deformed crosscap states as a function of subsystem size $\ell$. (Top left): Von Neumann entanglement entropy for three different values of $ \beta=e^{\mu/2}$, (Top right): R\'enyi-2 entanglement entropy for the same values of $ \beta$, (Bottom): R\'enyi-$n$ entanglement entropy for different indices $n=2,3,4$ for fixed $ \beta=0.1$.}
    \label{fig:crossasym:entropies_equilibrium}
\end{figure}
The static bipartite entanglement entropy of the fermionic crosscap states was first discussed in \cite{caetano2022crosscap} and its entanglement dynamics in~\cite{chalas2024quenchdynamicsentanglementcrosscap}. In this section we extend this analysis and examine the entanglement entropy and its  R\'enyi  generalisations for the charge deformed states. The $n^{\rm th}$ R\'enyi entanglement entropy is given by 
\begin{eqnarray}
    S^{(n)}[\rho_A]=\frac{1}{1-n}\log \tr [\rho_A^n],
\end{eqnarray}
from which the von-Neumann entropy is obtained in the replica limit $n
\to 1$. For a Gaussian state this can be expressed in terms of the covariance matrix as
\begin{eqnarray}\label{eq:crossasym:eigenvalue_renyi_ent}
    S^{(n)}[\rho_A]=\frac{1}{2(1-n)}\tr\log  \Big[\Big(\frac{1-\Gamma_A}{2}\Big)^n+\Big(\frac{1+\Gamma_A}{2}\Big)^n\Big]~.
\end{eqnarray}
Denoting the eigenvalues of $\Gamma_A$ by $\gamma_{m}$, where $m=1,\dots 2\ell$ this reduces to 
\begin{eqnarray}\label{eq:crossasym:renyi_ent_eigenvalues}
    S^{(n)}[\rho_A]&=&\frac{1}{2(1-n)} \sum_{m=1}^{2\ell} \log\Big[\Big(\frac{1-\gamma_m}{2}\Big)^n+\Big(\frac{1+\gamma_m}{2}\Big)^n\Big]~, 
\end{eqnarray}
and in the replica limit  we find the entanglement entropy,
\begin{eqnarray}
    S[\rho_A]
  = -\frac{1}{2}\sum_{m=1}^{2\ell} \Big[\Big(\frac{1-\gamma_m}{2}\Big)\log \Big(\frac{1-\gamma_m}{2}\Big)+\Big(\frac{1+\gamma_m}{2}\Big)\log \Big(\frac{1+\gamma_m}{2}\Big)\Big]~.
\end{eqnarray}
In the previous section we have determined these eigenvalues and inserting them above we find that the R\'enyi entropies are given by
\begin{eqnarray}\label{eq:crossasym:initial_renyi_result}
    S^{(n)}[\rho_A]={\rm min}(\ell,2L-\ell) s^{(n)}_ \mu
\end{eqnarray}
where 
\begin{eqnarray}\label{eq:crossasym:ent_density}
    s^{(n)}_ \mu=\frac{1}{1-n}\log[(1-\theta_ \mu)^n+\theta_ \mu^n]
\end{eqnarray}
with $s_ \mu\equiv \lim_{n\to 1}s^{(n)}_ \mu=-\theta_ \mu\log\theta_ \mu-(1-\theta_ \mu)\log(1-\theta_ \mu)$ in the replica limit. We plot these expressions in Figure \ref{fig:crossasym:entropies_equilibrium} for different values of $ \mu$ as a function of subsystem size. There, we see that the states $\ket{\mathcal{C}_\mu}$ exhibit similar properties to the $ \mu=0$ case, namely a Page curve like behaviour of the entanglement. Note, however, that as $| \mu|$ is increased the entanglement is suppressed so that for fixed system size it is monotonically decreasing as a function of $| \mu|$ becoming zero as $ \mu\to\pm\infty$ when $\ket{\mathcal{C}_ \mu}$ becomes a product state.  

This modification in the entanglement profile is clearly understood from the structure of $\rho_A$ given in~\eqref{eq:crossasym:reduced_density}. To see this we compute 
\begin{eqnarray}\nonumber
    \rho_A^n= \frac{1}{(1+e^\mu)^{|A_2|/2}} e^{\frac{1}{2}\sum_{x,y\in A_2}G^\beta_{x,y}c^
    \dag_x c^\dag_y} \prod_{x\in A_2}(1-c^\dag_xc_x)e^{-\frac{1}{2}\sum_{i,j\in A_2}G^\beta_{x,y}c_x c_y} \\\label{eq:crossasym:replicated_reduced_density}   
    \times \frac{e^{n \mu Q_{A_1}}}{(\tr [e^{ \mu Q_{A_1}}])^n}~.
\end{eqnarray}
Thus, only the mixed contribution in the second line depends on the Rényi index and, consequently, only this term contributes to the entanglement entropy. Taking the trace and recalling that $|A_1|=\min(\ell,2L-\ell)$, we recover Eq.~\eqref{eq:crossasym:initial_renyi_result}, which coincides with the thermodynamic Rényi entropy of the Gibbs state ${e^{ \mu Q_{A_1}}}/ {\tr [e^{ \mu Q_{A_1}}]}$. Accordingly, the entanglement entropy is maximal at $\mu=0$ and decreases as $|\mu|$ increases, reflecting the progressive bias towards a definite charge configuration.

\subsection{Entanglement Asymmetry }
\begin{figure}
    \centering
    \includegraphics[width=\linewidth]{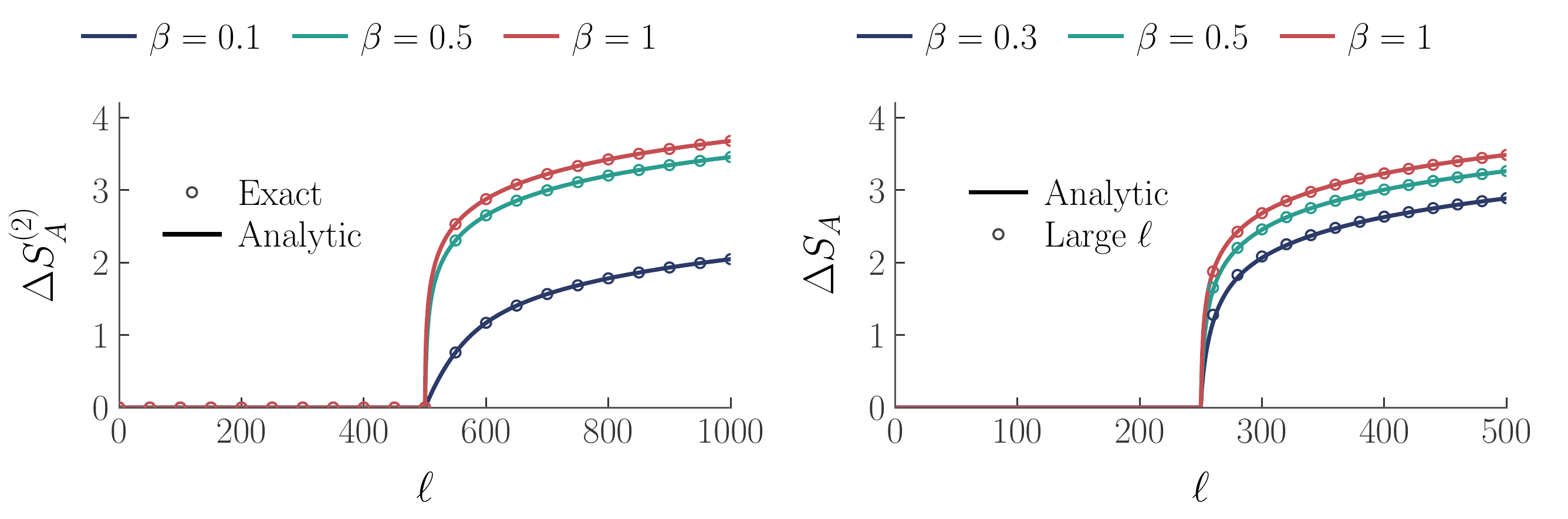}    
    % % \includegraphics[width=0.45\linewidth]{initia_state_renyi-2_asymm_3betas_2L=1000.pdf}
    % \includegraphics[width=0.49\linewidth]{vn_asym_initial_state_analytic_vs_large_ell.pdf}
    \caption{Entanglement asymmetries as function of subsystem size. (Left): R\'enyi-2 entanglement asymmetry versus subsystem size $\Delta S_A^{(2)}(\ell)$ for $ \beta=0.1,0.5,1$, exact results using equations~\eqref{eq:crossasym:def_charged_moments} and~\eqref{eq:crossasym:chraged_moment_asymmetry} (circles) compared with the analytical prediction of Eq.~\eqref{eq:crossasym:renyi_n_asymm} (solid line) for $2L=1000$. (Right): Von Neumann entanglement asymmetry $\Delta S_A(\ell)$ for  $\beta=0.3,0.5,1$, using the analytical  (solid lines) expressions~~\eqref{eq:crossasym:vn_asym_jq} and the large subsystem size limit in equation~\eqref{eq:crossasym:approx_asymmetry} (circles), in the case of $2L=500$.}
    \label{fig:crossasym:asymmetries_equilibrium}
\end{figure}
The entanglement asymmetry \cite{ares2023entanglement} is the relative entropy between the reduced density matrix and the symmetric state from which it is least distinguishable, $\bar\rho_{A,Q}$~\cite{gour2009measuring}. The form of $\bar\rho_{A,Q}$ means that $\Delta S$ can be written as the difference of entanglement entropies of the two states
\begin{eqnarray}\label{eq:crossasym:differnce_asymmetry}
    \Delta S[\rho_{A}]=S[\bar\rho_{A,Q}]-S[\rho_{A}]~,
\end{eqnarray}
which is much easier to treat. Although the entanglement asymmetry is a relative entropy, and therefore admits numerous R\'enyi versions,  the  difference form~\eqref{eq:crossasym:differnce_asymmetry} leads naturally to the following definition of the R\'enyi entanglement asymmetries 
\begin{eqnarray}
    \Delta S^{(n)}[\rho_{A}]=S^{(n)}[\bar\rho_{A,Q}]-S^{(n)}[\rho_{A}]~.
\end{eqnarray}
These have been regularly used as a proxy for the behaviour of $\Delta S$ which is recovered in the replica limit $n\to 1$. The most significant of the R\'enyi asymmetries is $n=2$ which has been experimentally measured in different platforms~\cite{ASYMM07joshi2024observing,Joshi2026}. While $\rho_A$ is a Gaussian state, its symmetrisation, $\bar\rho_{A,Q}$, can be very far from one~\cite{travaglino2026gaussianasymmetrymeasure} and so one cannot directly relate the eigenvalues of $\Gamma_A$ to $\Delta S$, presenting a challenge to analytic approaches. 

One route to calculate $\Delta S^{(n)}$ for a Gaussian state, however, is through the charged moments,  $ Z_{n}(\vec\lambda)$~\cite{ares2023entanglement,ASYMM06lackofrestoration}, which are defined as
\begin{eqnarray}
    Z_{n}(\vec\lambda)=\tr [\prod_{j=1}^n \rho_A e^{i (\lambda_j-\lambda_{j+1})Q_A}]
    \label{eq:crossasym:def_charged_moments}
\end{eqnarray}
where $\lambda_{n+1}=\lambda_1$. In terms of these we have 
\begin{eqnarray}\label{eq:crossasym:chraged_moment_asymmetry}
    \Delta S^{(n)}[\rho_A]=\frac{1}{1-n}\log \int_{-\pi}^\pi\prod_{j=1}^n \frac{{\rm d}\lambda_j}{2\pi}\frac{Z_n(\vec{\lambda})}{Z_n(\vec{0})} ~.
\end{eqnarray}
The quantity $Z_n(\vec\lambda)$ can then be expressed in terms of the covariance matrix as~\cite{ares2023entanglement,ASYMM06lackofrestoration}
\begin{equation}
\label{eq:crossasym:charged_mom_order_n_gaussian}
    Z_n(\vec{\lambda})=\sqrt{\det \left[\left(\frac{I-\Gamma_A}{2}\right)^n\left(I+\prod_{j=1}^nW_j\right)\right]},
\end{equation}
where $W_j=\left(I+\Gamma_A\right)\left(I-\Gamma_A\right)^{-1}e^{i(\lambda_{j}-\lambda_{j+1})q_A}$ (notice that, as usual, the eigenvalues of $\Gamma_A$ equal to $1$ do not cause any singularity in the complete determinant expression because they cancel with the prefactor having  zeros of the same order exactly in $1$), and $q_A$ is a matrix with elements $(q_A)_{2m,2m}=+1$ and $(q_A)_{2m-1,2m-1}=-1$, with $m=1,\cdots,\ell$. Although exact, this expression is quite cumbersome to analyse, especially for higher $n$. 

Alternatively, since we have access to the reduced density matrix, we can try to calculate $Z_n(\vec{\lambda})$ directly using~\eqref{eq:crossasym:reduced_density}. We find that 
\begin{eqnarray}\label{eq:crossasym:initia_charegd_moment}
    Z_n(\vec{\lambda})=\prod_{j=1}[1-\theta_ \mu +e^{2i(\lambda_j-\lambda_{j+1})}\theta_ \mu]^{|A_2|/2} \big[(1-\theta_ \mu)^n+\theta_ \mu^n\big]^{{|A_1|}}~.
\end{eqnarray}
Here, one can surmise that the $\lambda_j$ dependent terms come from the first three terms  of~\eqref{eq:crossasym:reduced_density}, i.e. the pure part, where both antipodally correlated sites are included, whereas the $\lambda_j$ independent but $n$ dependent terms comes from the rightmost term which is the mixed one, corresponding to antipodal site-pairs where one site is traced out of the subsystem of interest.

Using~\eqref{eq:crossasym:initia_charegd_moment} in~\eqref{eq:crossasym:chraged_moment_asymmetry} and performing some changes of variables~\cite{rylands2024microscopic} we find that
\begin{eqnarray}
    \Delta S^{(n)}[\rho_A]=\frac{1}{1-n}\log\left[\sum_{q=0}^{|A_2| } J(q)^n\right]
    \label{eq:crossasym:renyi_n_asymm}
\end{eqnarray}
and in the replica limit
\begin{eqnarray}
    \Delta S[\rho_A]=-\sum_{q=0}^{|A_2| } J(q)\log J(q)~,
    \label{eq:crossasym:vn_asym_jq}
\end{eqnarray}
where we have introduced 
\begin{eqnarray}
     J(q)&=&\int_{-\pi}^\pi\frac{{\rm d}\lambda}{2\pi}\,\big(1-\theta_ \mu+e^{2i\lambda}\theta_ \mu\big)^{|A_2| /2}e^{-i\lambda q }\\
     &=&\begin{cases}
         \binom{|A_2| /2}{q/2}\theta_ \mu^{q/2}(1-\theta_ \mu)^{(|A_2| -q)/2}~&~\text{for $q$ even}\\
         0~&~\text{for $q$ odd}
     \end{cases}
\end{eqnarray}
These expressions are similar to what would be expected upon calculating the entanglement asymmetry for a full system in a pure state rather than in a subsystem which is described by a mixed reduced density matrix. In our case we have a factorisation of the density matrix into pure and mixed parts, only the former of which contributes to the symmetry breaking. Indeed, the quantity $J(q)$ can be interpreted as the  probability distribution for $Q_{A_2}$, the charge within the region $A_2$. 

For $|A_2|>1$ and away from $ \mu\to\pm\infty$ these can be very well approximated by
\begin{eqnarray}
    \Delta S^{(n)}[\rho_A]\simeq \frac{1}{2}\log[ \pi n^{\frac{1}{n-1}} |A_2|\theta_ \mu(1-\theta_ \mu)]
\end{eqnarray}
and 
\begin{eqnarray}\label{eq:crossasym:approx_asymmetry}
    \Delta S[\rho_A]\simeq \frac{1}{2}\log[ \pi e |A_2| \theta_ \mu(1-\theta_ \mu)]~.
\end{eqnarray}
We plot these quantities for different $ \mu$  and $n$ as a function of subsystem size in Figure \ref{fig:crossasym:asymmetries_equilibrium}. For $\ell\leq L$ the subsystem is symmetric, with a vanishing $\Delta S$. As the subsystem size is increased, however, the asymmetry monotonically increases.
We previously observed that the leading behaviour of the entanglement entropy  of the crosscap states is identical to the Page curve for Haar random states. 
While the entanglement asymmetry has some features which are reminiscent of the Page curve for asymmetry \cite{ampc-23}, there are some important distinctions. In fact in both cases the state is symmetric for $\ell\leq L$.
Conversely at $\ell=L$ Haar random states exhibit a sharp jump~\cite{ampc-23,Russotto_2025}, while charge-deformed crosscap states have continuous asymmetry also for large subsystems.
Another important distinction to be made between the two cases, is that the crosscap state is a single pure quantum state, whereas the results of the Haar random states regard averages over an ensemble of random states in the thermodynamic limit.

 \subsection{Symmetry resolution}
We now examine the interplay between  entanglement, symmetry and asymmetry in the deformed crosscap states.  We start with subsystems which are less than half the system size,  $\ell\leq L$. In this case the subsystem $A$ contains no antipodal points and so
\begin{eqnarray}
    \rho_{A}=\frac{e^{ \mu Q_{A}}}{\tr [e^{ \mu Q_{A}}]}~.
\end{eqnarray}
This is a symmetric state and therefore admits a decomposition in terms of the number and configuration entropy~\eqref{eq:crossasym:symm_resolution}. These can be calculated explicitly whereupon we find that the number entropy is 
\begin{eqnarray}
    S_{\rm num}[\rho_A]=-\sum_{q=0}^{|A_1|}p_<(q)\log p_<(q)~,\\
    p_<(q)=\binom{|A_1|}{q}(\theta_ \mu)^q(1-\theta_ \mu)^{|A_1|-q}~,
\end{eqnarray}
 Here, we have denoted by $p_<(q)$ the charge probability distributions for subsystems of this size and, since $\ell\leq L$, we have that $|A_1|=\ell$.  The configuration entropy is then given by
\begin{eqnarray}
S_{\rm conf}[\rho_A]=\sum_{q=0}^\ell p_<(q) S[\rho_{A,q}]~,
\end{eqnarray}
with the symmetry resolved entropy being
\begin{eqnarray}
S[\rho_{A,q}]=\log\binom{|A_1|}{q}~.
\end{eqnarray}
This latter result indicates that after a projective measurement of charge, the subsystem is maximally mixed within that charge sector.  It can then be explicitly checked using properties of the binomial distributions that these resum to give precisely $S_{\rm num}[\rho_A]+S_{\rm conf}[\rho_A]=\ell s_ \mu$. 

For larger subsystems, the entanglement decreases taking on the same value as in complement of $A$, see Figure \ref{fig:crossasym:entropies_equilibrium}, however, the number of accessible charge sectors increases and so the interplay between charge and entanglement must be modified. Indeed, as we have just seen, however, the subsystem for $\ell>L$ is not symmetric and so the entanglement entropy, instead admits the decomposition~\eqref{eq:crossasym:asymm_resolution} in terms of the number and configuration parts as well as the asymmetry. The number entropy is
\begin{eqnarray}
    S_{\rm num}[\rho_A]=-\sum_{q=0}^{\ell}p_>(q)\log p_>(q)~,\\
    p_>(q)=\sum_{q'=0}^qp_<(q')J(q-q'),
\end{eqnarray}
where we see that the charge probability distribution for $A$ is a simple convolution of the disjoint distributions of charge within $|A_1|$ and $|A_2|$. %We note that the form of $J(q)$, and the fact that $p_{<}(q)=0$ for $q>|A_1|$, dictate that not all values of the charge can be measured inside $A$.  

The configuration entropy can also be determined. It is complicated by the change in the allowed charge measurement outcomes but then simplified by the factorisation of the reduced density matrix.  Indeed, 
although the reduced density matrix does not generally commute with the total subsystem charge $Q_A$, it does commute with $Q_{A_1}$, as follows from its factorized structure in Eq. \eqref{eq:crossasym:reduced_density}. Consequently, the charge-projected state $\rho_{A,q}$ is block diagonal in the eigenbasis of $Q_{A_1}$.
%while the projected reduced density matrix, $\rho_{A,q}$ is symmetric with respect to $Q_A$, it is also symmetric with respect to $Q_{A_1}$, as seen in \eqref{eq:crossasym:reduced_density}. Thus, $\rho_{A,q}$ is block diagonal in the basis of $Q_{A_1}$ and so $S[\rho_{A,q}]$ can be further resolved with respect to $Q_{A_1}$. 
The result is that 
\begin{eqnarray}\label{eq:crossasym:conf_ent}
     S[\rho_{A,q}]=-\sum_{q'=0}^{|A_1|}p(q'|q)\log p(q'|q)+\sum_{q'=0}^{|A_1|}p(q'|q)\log \binom{|A_1|}{q'}.
\end{eqnarray}
where here 
\begin{eqnarray}
    p(q'|q)=\frac{p_<(q')J(q-q')}{p_>(q)}
\end{eqnarray}
is the probability that a measurement of $Q_{A_1}$ returns a value of $q'$ given that a measurement of  $Q_A$ has returned a value $q$.  The first term in \eqref{eq:crossasym:conf_ent} is the contribution from charge fluctuations in $A_1$. The second term is the configurational entropy and can be obtained by noting that only the mixed term can contribute to the entropy. 
Despite the complicated nature of both the number and configuration entropies one can confirm that they sum up to give $(2L-\ell)s_ \mu+\Delta S[\rho_A]$. Thus, the increase in both number and configuration entropy which occurs due to the larger number of accessible charge sectors is cancelled by the fact that the state is no longer symmetric through the non-zero entanglement asymmetry.

\section{Quench dynamics}\label{sec:crossasym:quecnh}

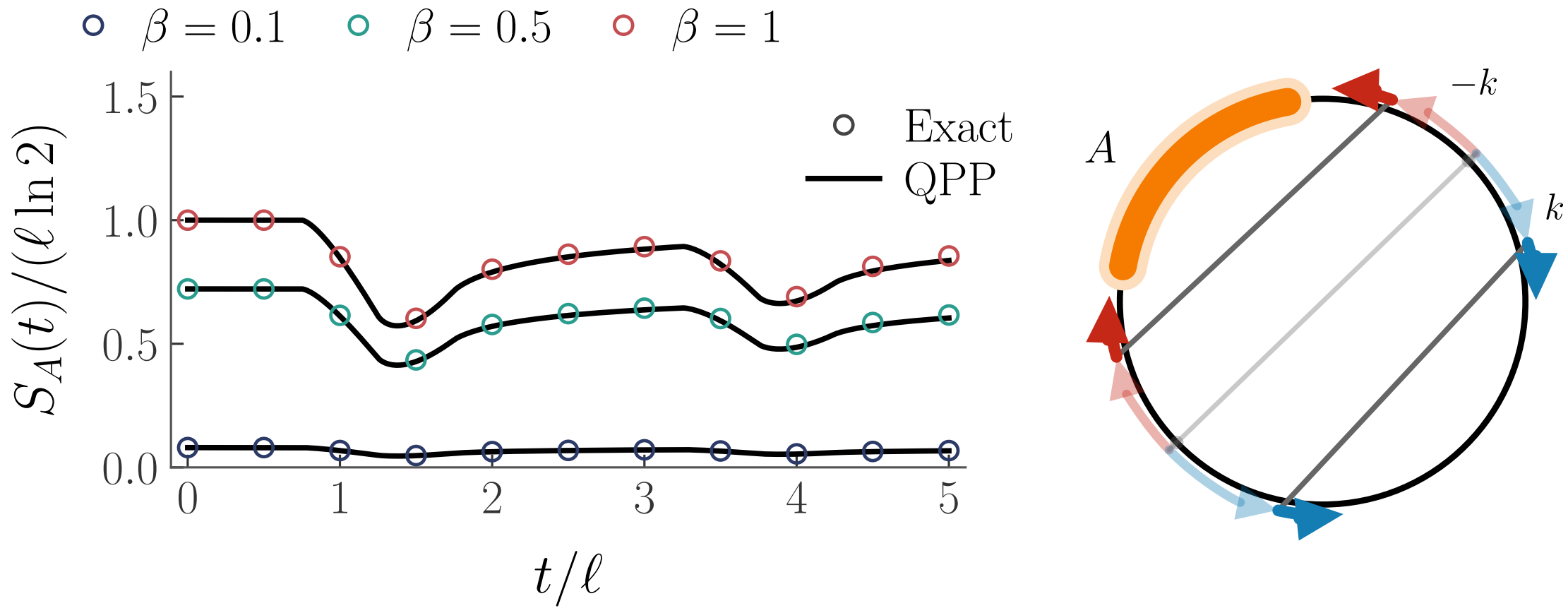
\begin{figure}[t]
    \centering

    % ========================================================
    % LEFT PANEL: Numerical / QPP plot
    % ========================================================
    \begin{minipage}[c]{0.66\linewidth}
        \centering
        \includegraphics[width=\linewidth]{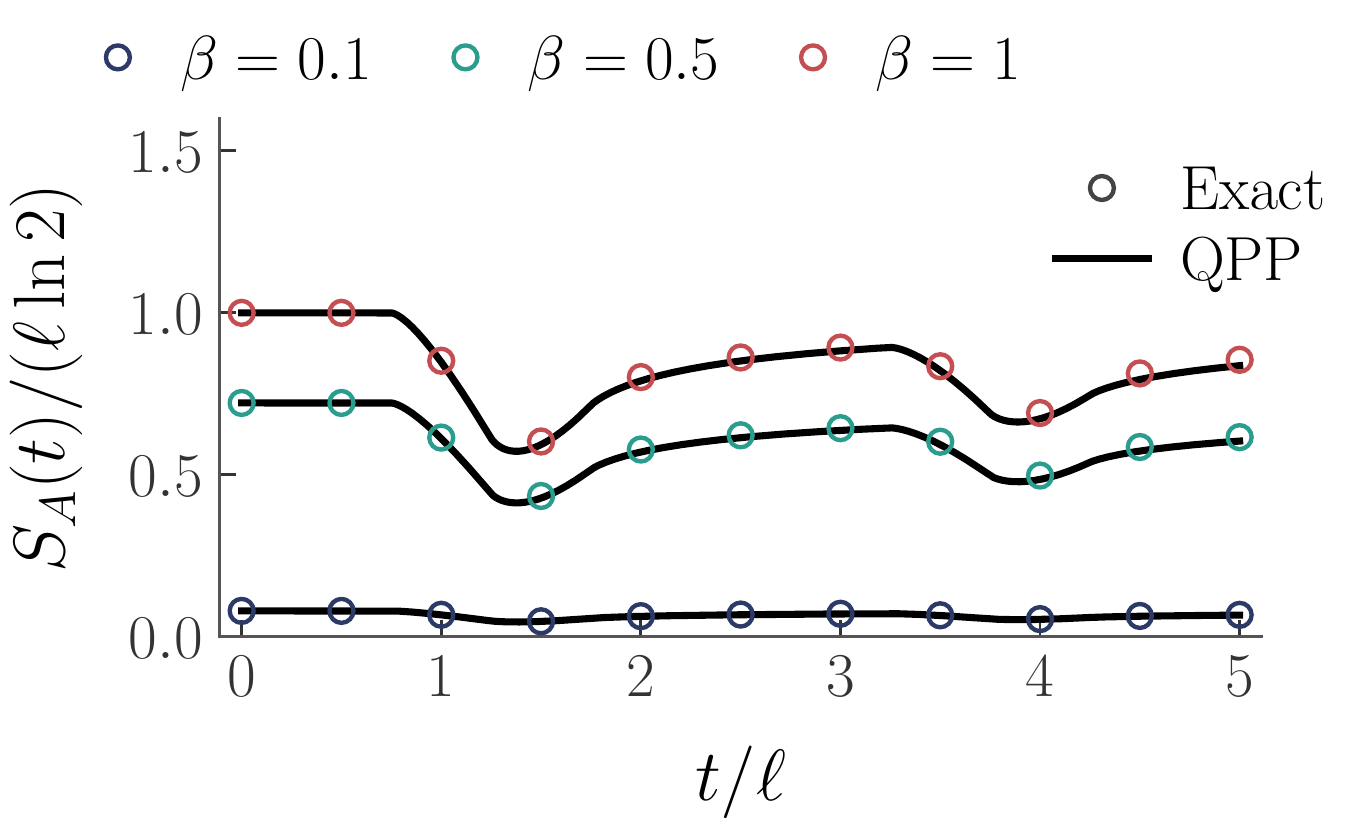}
    \end{minipage}%
    \hfill%
    % ========================================================
    % RIGHT PANEL: TikZ quasiparticle schematic
    % ========================================================
    \begin{minipage}[c]{0.32\linewidth}
        \centering

        \definecolor{qpblue}{HTML}{147DB3}
        \definecolor{qpred}{HTML}{C62817}
        \definecolor{suborange}{HTML}{F57C00}
        \definecolor{guidegray}{HTML}{666666}

        \resizebox{\linewidth}{!}{%
        \begin{tikzpicture}[
            line cap=round,
            line join=round,
            scale=0.58,
            every node/.style={transform shape}
        ]

        % ====================================================
        % Parameters
        % ====================================================
        \def\R{2.30}
        \def\Router{2.42}

        \def\Astart{100}
        \def\Aend{170}

        \def\QPa{44}
        \def\QPb{224}

        % First snapshot endpoints
        \def\redUp{71}
        \def\blueUp{16}
        \def\redDown{195}
        \def\blueDown{258}

        % Second snapshot endpoints
        \def\redUpLater{88}
        \def\blueUpLater{0}
        \def\redDownLater{178}
        \def\blueDownLater{276}

        % ====================================================
        % Main ring
        % ====================================================
        \draw[
            black,
            line width=1.15pt
        ]
        (0,0) circle (\R);

        % ====================================================
        % Subsystem A
        % ====================================================
        \draw[
            suborange!25,
            line width=9pt
        ]
        (\Astart:\R)
        arc[
            start angle=\Astart,
            end angle=\Aend,
            radius=\R
        ];

        \draw[
            suborange,
            line width=5.2pt
        ]
        (\Astart:\R)
        arc[
            start angle=\Astart,
            end angle=\Aend,
            radius=\R
        ];

        \node[
            font=\Large,
            anchor=east
        ]
        at (-2.20,1.75)
        {$A$};

        % ====================================================
        % INITIAL CONFIGURATION — FADED
        % ====================================================

        \draw[
            guidegray,
            line width=0.9pt,
            opacity=0.35
        ]
        (\QPa:\Router) -- (\QPb:\Router);

        \fill[guidegray,opacity=0.35]
        (\QPa:\R) circle (2.2pt);

        \fill[guidegray,opacity=0.35]
        (\QPb:\R) circle (2.2pt);

        % Upper red
        \draw[
            qpred,
            line width=1.7pt,
            opacity=0.35,
            -{Triangle[length=2.7mm,width=2.2mm]}
        ]
        (\QPa:\Router)
        arc[
            start angle=\QPa,
            end angle=\redUp,
            radius=\Router
        ];

        % Upper blue
        \draw[
            qpblue,
            line width=1.7pt,
            opacity=0.35,
            -{Triangle[length=2.7mm,width=2.2mm]}
        ]
        (\QPa:\Router)
        arc[
            start angle=\QPa,
            end angle=\blueUp,
            radius=\Router
        ];

        % Lower red
        \draw[
            qpred,
            line width=1.7pt,
            opacity=0.35,
            -{Triangle[length=2.7mm,width=2.2mm]}
        ]
        (\QPb:\Router)
        arc[
            start angle=\QPb,
            end angle=\redDown,
            radius=\Router
        ];

        % Lower blue
        \draw[
            qpblue,
            line width=1.7pt,
            opacity=0.35,
            -{Triangle[length=2.7mm,width=2.2mm]}
        ]
        (\QPb:\Router)
        arc[
            start angle=\QPb,
            end angle=\blueDown,
            radius=\Router
        ];

        % ====================================================
        % LATER CONFIGURATION — SOLID
        % ====================================================

        % Red pair: connect bases
        \draw[
            guidegray,
            line width=1.0pt
        ]
        (\redUp:\Router) -- (\redDown:\Router);

        % Blue pair: connect bases
        \draw[
            guidegray,
            line width=1.0pt
        ]
        (\blueUp:\Router) -- (\blueDown:\Router);

        % Upper red
        \draw[
            qpred,
            line width=2.2pt,
            -{Triangle[length=3.2mm,width=2.6mm]}
        ]
        (\redUp:\Router)
        arc[
            start angle=\redUp,
            end angle=\redUpLater,
            radius=\Router
        ];

        % Upper blue
        \draw[
            qpblue,
            line width=2.2pt,
            -{Triangle[length=3.2mm,width=2.6mm]}
        ]
        (\blueUp:\Router)
        arc[
            start angle=\blueUp,
            end angle=\blueUpLater,
            radius=\Router
        ];

        % Lower red
        \draw[
            qpred,
            line width=2.2pt,
            -{Triangle[length=3.2mm,width=2.6mm]}
        ]
        (\redDown:\Router)
        arc[
            start angle=\redDown,
            end angle=\redDownLater,
            radius=\Router
        ];

        % Lower blue
        \draw[
            qpblue,
            line width=2.2pt,
            -{Triangle[length=3.2mm,width=2.6mm]}
        ]
        (\blueDown:\Router)
        arc[
            start angle=\blueDown,
            end angle=\blueDownLater,
            radius=\Router
        ];

        % ====================================================
        % Momentum labels
        % ====================================================

        \node[
            font=\large,
            anchor=south west
        ]
        at (1.30,2.18)
        {$-k$};

        \node[
            font=\large,
            anchor=west
        ]
        at (2.38,1.08)
        {$k$};

        \end{tikzpicture}%
        }

    \end{minipage}

    % ========================================================
    % Caption
    % ========================================================
    \caption{
        Left: The entanglement entropy as a function of time
        for $\beta=0.1,0.5,1$, for a chain of size
        $\mathcal{L}=400$ and a subsystem of length $\ell=80$.
        The solid lines are the predictions of the quasiparticle
        picture~\eqref{eq:crossasym:ent_ent_qpp},
        while the symbols are the exact numerical results
        obtained using
        \eqref{eq:crossasym:renyi_ent_eigenvalues}.
        Right: Schematic representation of the motion of
        initially antipodally entangled quasiparticle pairs.
        The red and blue quasiparticles propagate in opposite
        directions, with the faded configuration indicating
        their earlier positions.
    }

    \label{fig:crossasym:entanglement_qpp}

\end{figure}

Having studied the initial-state properties of the deformed crosscap states in the previous section we now examine the quench dynamics from these states. Taking $\ket{\mathcal{C}_ \mu}$ as the initial state, we unitarily evolve the system under the tight binding Hamiltonian,
\begin{equation}
    H=-\sum_{x=1}^{2L} [c^\dagger_x c_{x+1}+c^{\dagger}_{x+1}c_{x}]
=\sum_{m=1}^{2L}\epsilon(k_m)c^\dag_{k_m}c_{k_m}.
\end{equation}
In the second line we have switched to Fourier space and introduced  
\begin{equation}
\label{eq:crossasym:sec2:Fourier-def}
   c_{k_m}=\frac{1}{\sqrt{2L}}\sum_{x=1}^{2L}c_{x}e^{-ik_m x};\qquad     k_m=\frac{\pi}{L}\left(m-\frac{1}{2}\right);\qquad m=1,\cdots, 2L
\end{equation}
and $\epsilon(k)=-2\cos(k)$.  
The time evolved state, $\ket{\mathcal{C}_ \mu(t)}=e^{-i Ht}\ket{\mathcal{C}_ \mu}$,  remains Gaussian and so all subsystem properties can be understood using the time evolved, restricted covariance matrix, $\Gamma_A(t)$ which we express in Fourier space as
 \begin{equation}
 \label{eq:crossasym:sec2:time_evolved_corrmatr_gamma_crosscaps}
[ \Gamma(t)]_{x,y}
     = \frac{1}{L}\sum_{m=1}^{2L}e^{ik_m(x-y)} \begin{pmatrix}
~  \theta_ \mu-\frac{1}{2}  & \sqrt{\theta_ \mu(1-\theta_ \mu)} e^{ik_mL-2 i t\epsilon(k_m)}\\
    \sqrt{\theta_ \mu(1-\theta_ \mu)} e^{-ik_mL+2 it\epsilon(k_m)} & \frac{1}{2}-\theta_ \mu
\end{pmatrix}.
\end{equation}
 Once again, the eigenvalues of the covariance matrix determine the dynamics of the R\'enyi entanglement entropies through~\eqref{eq:crossasym:eigenvalue_renyi_ent} which gives a means to determine the entanglement, at least numerically. Nevertheless, the entanglement dynamics from undeformed crosscap states in free fermion models has been studied previously~\cite{chalas2024quenchdynamicsentanglementcrosscap} and an effective quasiparticle picture of the dynamics determined. The quasiparticle picture for the charge deformed states, as we will see, is underpinned by the same quasiparticle structure with some modifications, and so we briefly review it in the next section.

\subsection{Quasiparticle picture for entanglement}
The quasiparticle picture is an effective, ballistic scale and semi-classical description of quench dynamics in integrable models~\cite{calabrese2005evolution,alba2017entanglement} which has now been generalised to numerous quantities beyond entanglement~\cite{alba2019quantum,ares2023entanglement,groha_fcs_tfim_2018,alba2019scrambling,rottoli2024entanglement,Travaglino_2025}. The starting point for the entanglement entropies is to expand them as a power series in traces of powers of the covariance matrix which can then be evaluated by a stationary phase approximation in the limit of large $\ell,t$ but with $t/\ell$ fixed~\cite{fagotti2008evolution,fagotti2010entanglement,calabrese2012quantum,chalas2025entanglementevolutionentangledmultipodal}. For $\ell<L$ the result for the charge deformed crosscap states is 
\begin{eqnarray}\label{eq:crossasym:ent_ent_qpp}
    S^{(n)}[\rho_A(t)]&=&s^{(n)}_ \mu\ell_{\rm s}(t),\\
    \ell_{\rm s}(t)&=& \int_{-\pi}^\pi \frac{{\rm d}k}{2\pi}[ \ell-{\rm max}(0,\ell -|2 \tau_k |v(k)|-L|)]
\end{eqnarray}
where $s^{(n)}_ \mu$ is given in~\eqref{eq:crossasym:ent_density}, $v(k)=2\sin(k)$ and $\tau_k=t~{\rm mod}[L/|v(k)|]$. The reciprocity of entanglement entropy then gives us the entanglement dynamics for subsystems with $\ell>L$.
The above expression has a transparent physical interpretation in terms of the quasiparticle picture, see also the right panel of Figure \ref{fig:crossasym:entanglement_qpp}. In particular, the quench creates correlated pairs of quasiparticles which propagate throughout the system thereby creating entanglement between different regions. In a quench from a lowly entangled state, correlations only exist between pairs which emanate from the same point in space and have the same magnitude of velocity $|v(k)|$. In the crosscap quench, correlated quasiparticles  are emitted from antipodal points and propagate toward each other but with velocity of the same magnitude $| v(k)|$. The factor $\ell_{\rm s}(t)$, counts the number of correlated pairs which are shared between $A$ and $\bar A$ as a function of time with the time variable $\tau_k$ taking into account the finite size of the system \cite{modak2020entanglement,Lagnese_2022,chalas2024quenchdynamicsentanglementcrosscap}. Each shared pair contributes $s^{(n)}_ \mu$ to the value of the R\'enyi entanglement entropy. This expression reduces to the previously derived one for $ \mu=0$. 

We compare this expression to the exact result obtained by numerically determining the eigenvalues of $\Gamma_A(t)$ and~\eqref{eq:crossasym:eigenvalue_renyi_ent} in the left panel of Figure \ref{fig:crossasym:entanglement_qpp} finding excellent agreement at all times.  The qualitative behaviour is the same for all finite $ \mu$, the entanglement is constant until a time where correlated pairs can become completely contained inside $A$ whereupon they result in a decrease in entanglement. After a period of decrease,   the entanglement starts increasing as a result of correlated pairs becoming shared again.  Subsequent oscillations  can all be explained by the propagation of correlated pairs around system~\cite{chalas2024quenchdynamicsentanglementcrosscap}. For increasing $| \mu|$ the overall value of the entanglement is decreased, in line with the same observation in the initial state.

\section{Asymmetry dynamics }\label{sec:crossasym:asymm}

\begin{figure}
    \centering
    \includegraphics[width=\linewidth]{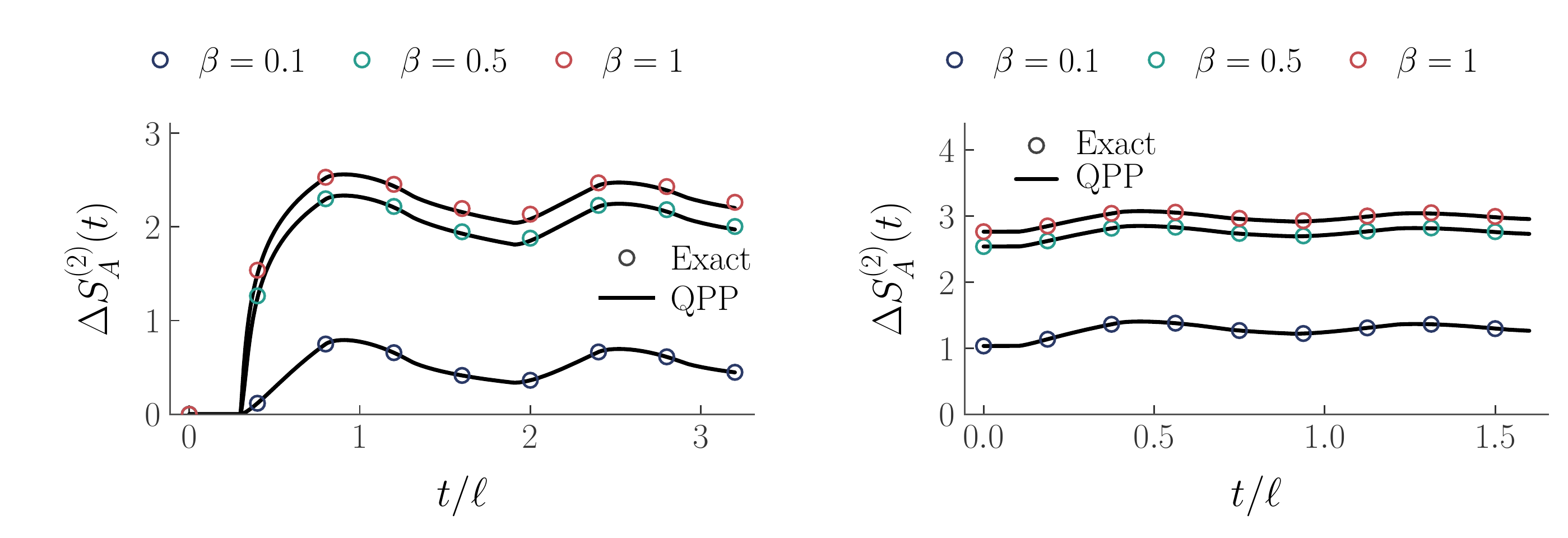}
    \caption{R\'enyi-2 entanglement asymmetry $\Delta S_A^{(2)}(t)$ as a function of time, for different values of $\beta=0.1,0.5,1$ for a fermionic system of size $\mathcal{L}=640$.  Left: Time evolution for a subsystem of $\ell<L$, in particular $\ell=200$ starting from a symmetric subsystem. Right: for a subsystem $\ell>L$, $\ell=400$ starting from a symmetry-breaking subsystem. The solid lines are the predictions of the quasiparticle picture \eqref{eq:crossasym:time_evolved_renyi_asymm} while the symbols are the exact numerical result obtained from~\eqref{eq:crossasym:charged_mom_order_n_gaussian}}.
    \label{fig:crossasym:renyi2_asymmetry_dynamics}
\end{figure}

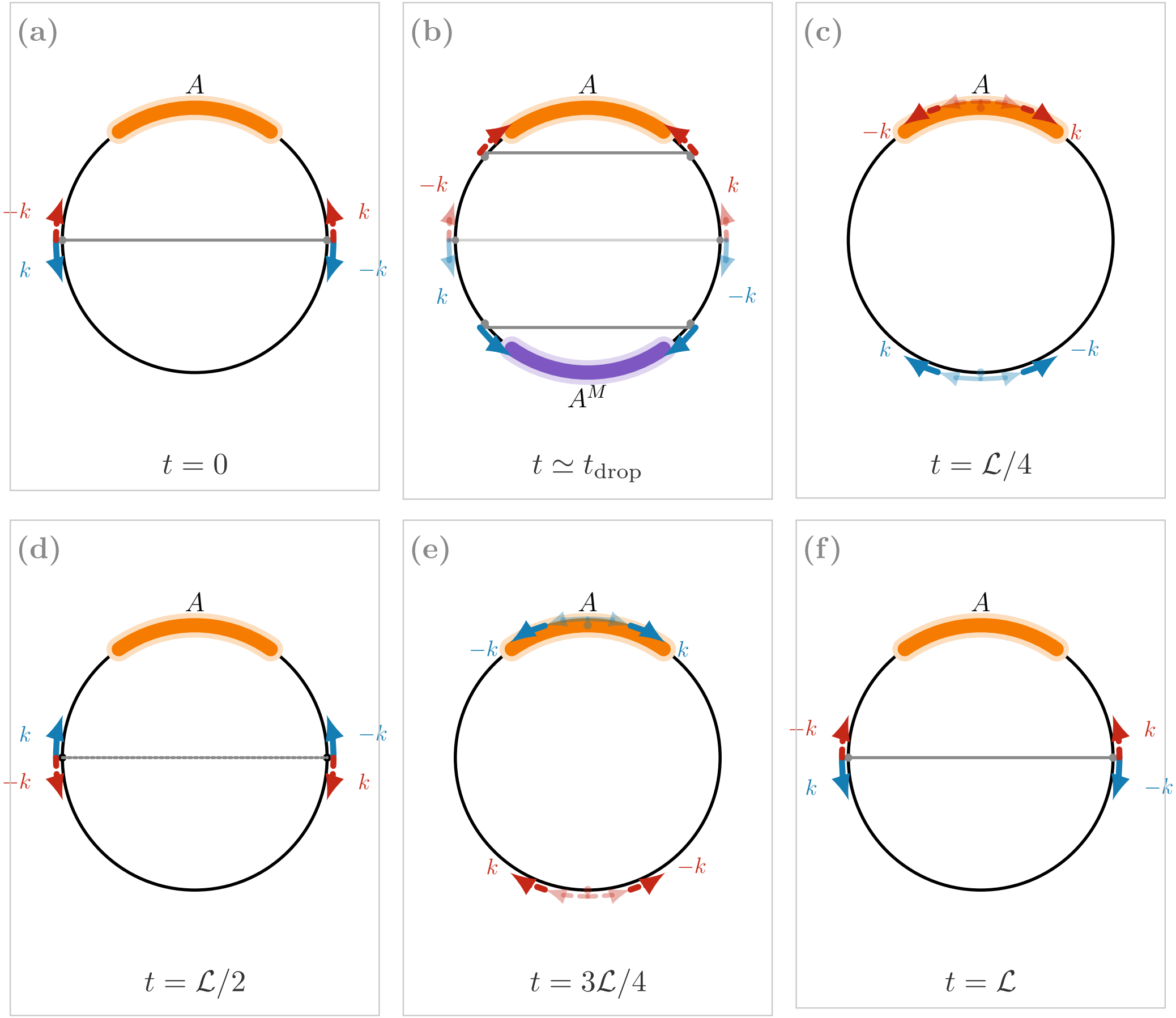
\begin{figure*}[t]
\centering

\definecolor{qpblue}{HTML}{147DB3}
\definecolor{qpred}{HTML}{C62817}
\definecolor{suborange}{HTML}{F57C00}
\definecolor{mirrorviolet}{HTML}{7E57C2}
\definecolor{guidegray}{HTML}{8A8A8A}

% ============================================================
% ORIGINAL SLIDE PANEL (b)
% ============================================================

\def\CCpanelB{%
\begin{tikzpicture}[
    line cap=round,
    line join=round,
    scale=0.68,
    every node/.style={transform shape}
]

\path[use as bounding box]
(-3.3,-3.45) rectangle (3.3,3.30);

\def\R{2.50}
\def\Router{2.62}
\def\Astart{55}
\def\Aend{125}
\def\Mstart{235}
\def\Mend{305}
\def\delta{16}

\def\StartLeft{180}
\def\StartRight{0}

\pgfmathsetmacro{\RedLeft}{\Aend+\delta}
\pgfmathsetmacro{\RedRight}{\Astart-\delta}
\pgfmathsetmacro{\BlueLeft}{180+\RedRight}
\pgfmathsetmacro{\BlueRight}{\RedLeft-180}

% Main circle
\draw[black,line width=1.15pt]
(0,0) circle (\R);

% Subsystem A
\draw[suborange!25,line width=9pt]
(\Astart:\R)
arc[start angle=\Astart,end angle=\Aend,radius=\R];

\draw[suborange,line width=5.2pt]
(\Astart:\R)
arc[start angle=\Astart,end angle=\Aend,radius=\R];

\node[font=\Large,anchor=south]
at (0,2.63) {$A$};

% Mirror subsystem
\draw[mirrorviolet!25,line width=9pt]
(\Mstart:\R)
arc[start angle=\Mstart,end angle=\Mend,radius=\R];

\draw[mirrorviolet,line width=5.2pt]
(\Mstart:\R)
arc[start angle=\Mstart,end angle=\Mend,radius=\R];

\node[font=\Large,anchor=north]
at (0,-2.6) {$A^M$};

% Initial positions
\draw[guidegray,line width=1pt,opacity=0.40]
(\StartLeft:\Router) -- (\StartRight:\Router);

\fill[guidegray] (\StartLeft:\R) circle (2pt);
\fill[guidegray] (\StartRight:\R) circle (2pt);

% Earlier red propagation
\draw[qpred,dashed,line width=1.8pt,opacity=0.45,-latex]
(\StartLeft:\Router)
arc[start angle=180,end angle=164,radius=\Router];

\draw[qpred,dashed,line width=1.8pt,opacity=0.45,-latex]
(\StartRight:\Router)
arc[start angle=0,end angle=16,radius=\Router];

% Earlier blue propagation
\draw[qpblue,line width=1.8pt,opacity=0.45,-latex]
(\StartLeft:\Router)
arc[start angle=180,end angle=196,radius=\Router];

\draw[qpblue,line width=1.8pt,opacity=0.45,-latex]
(\StartRight:\Router)
arc[start angle=0,end angle=-16,radius=\Router];

% Current pair connections
\draw[guidegray,line width=1.1pt]
(\RedLeft:\Router) -- (\RedRight:\Router);

\draw[guidegray,line width=1.1pt]
(\BlueLeft:\Router) -- (\BlueRight:\Router);

\fill[guidegray] (\RedLeft:\R) circle (2.2pt);
\fill[guidegray] (\RedRight:\R) circle (2.2pt);
\fill[guidegray] (\BlueLeft:\R) circle (2.2pt);
\fill[guidegray] (\BlueRight:\R) circle (2.2pt);

% Red propagation
\draw[qpred,dashed,line width=2.2pt,-latex]
(\RedLeft:\Router)
arc[start angle=\RedLeft,
    end angle={\Aend-2},radius=\Router];

\draw[qpred,dashed,line width=2.2pt,-latex]
(\RedRight:\Router)
arc[start angle=\RedRight,
    end angle={\Astart+2},radius=\Router];

% Blue propagation
\draw[qpblue,line width=2.2pt,-latex]
(\BlueLeft:\Router)
arc[start angle=\BlueLeft,
    end angle={\Mstart+2},radius=\Router];

\draw[qpblue,line width=2.2pt,-latex]
(\BlueRight:\Router)
arc[start angle=\BlueRight,
    end angle=-57,radius=\Router];

% Momentum labels
\node[qpred,font=\large,anchor=east]
at (-2.5,1.05) {$-k$};

\node[qpred,font=\large,anchor=west]
at (2.5,1.05) {$k$};

\node[qpblue,font=\large,anchor=east]
at (-2.5,-1.05) {$k$};

\node[qpblue,font=\large,anchor=west]
at (2.5,-1.05) {$-k$};

\end{tikzpicture}%
}

% ============================================================
% ORIGINAL SLIDE PANEL (c)
% ============================================================

\def\CCpanelC{%
\begin{tikzpicture}[
    line cap=round,
    line join=round,
    scale=0.68,
    every node/.style={transform shape}
]

\path[use as bounding box]
(-3.3,-3.45) rectangle (3.3,3.30);

\def\R{2.50}
\def\Router{2.62}
\def\Astart{55}
\def\Aend{125}

\def\meetTop{90}
\def\meetBottom{270}
\def\dold{18}
\def\dnew{34}

% Main circle
\draw[black,line width=1.15pt]
(0,0) circle (\R);

% Subsystem A
\draw[suborange!25,line width=9pt]
(\Astart:\R)
arc[start angle=\Astart,end angle=\Aend,radius=\R];

\draw[suborange,line width=5.2pt]
(\Astart:\R)
arc[start angle=\Astart,end angle=\Aend,radius=\R];

\node[font=\Large,anchor=south]
at (0,2.63) {$A$};

% Top red pair
\fill[qpred,opacity=0.35]
(\meetTop:\R) circle (2.2pt);

% Earlier motion
\draw[qpred,dashed,line width=1.7pt,opacity=0.35,-latex]
(\meetTop:\Router)
arc[start angle=\meetTop,
    end angle={\meetTop+\dold},radius=\Router];

\draw[qpred,dashed,line width=1.7pt,opacity=0.35,-latex]
(\meetTop:\Router)
arc[start angle=\meetTop,
    end angle={\meetTop-\dold},radius=\Router];

% Later motion
\draw[qpred,dashed,line width=2.2pt,-latex]
({\meetTop+\dold}:\Router)
arc[start angle={\meetTop+\dold},
    end angle={\meetTop+\dnew},radius=\Router];

\draw[qpred,dashed,line width=2.2pt,-latex]
({\meetTop-\dold}:\Router)
arc[start angle={\meetTop-\dold},
    end angle={\meetTop-\dnew},radius=\Router];

% Bottom blue pair
\fill[qpblue,opacity=0.35]
(\meetBottom:\R) circle (2.2pt);

% Earlier motion
\draw[qpblue,line width=1.7pt,opacity=0.35,-latex]
(\meetBottom:\Router)
arc[start angle=\meetBottom,
    end angle={\meetBottom-\dold},radius=\Router];

\draw[qpblue,line width=1.7pt,opacity=0.35,-latex]
(\meetBottom:\Router)
arc[start angle=\meetBottom,
    end angle={\meetBottom+\dold},radius=\Router];

% Later motion
\draw[qpblue,line width=2.2pt,-latex]
({\meetBottom-\dold}:\Router)
arc[start angle={\meetBottom-\dold},
    end angle={\meetBottom-\dnew},radius=\Router];

\draw[qpblue,line width=2.2pt,-latex]
({\meetBottom+\dold}:\Router)
arc[start angle={\meetBottom+\dold},
    end angle={\meetBottom+\dnew},radius=\Router];

% Momentum labels
\node[qpred,font=\large,anchor=east]
at (-1.55,2.05) {$-k$};

\node[qpred,font=\large,anchor=west]
at (1.55,2.05) {$k$};

\node[qpblue,font=\large,anchor=east]
at (-1.55,-2.05) {$k$};

\node[qpblue,font=\large,anchor=west]
at (1.55,-2.05) {$-k$};

\end{tikzpicture}%
}

% ============================================================
% ORIGINAL SLIDE PANEL (d)
% ============================================================

\def\CCpanelD{%
\begin{tikzpicture}[
    line cap=round,
    line join=round,
    scale=0.68,
    every node/.style={transform shape}
]

\path[use as bounding box]
(-3.3,-3.45) rectangle (3.3,3.30);

\def\R{2.50}
\def\Router{2.62}
\def\Astart{55}
\def\Aend{125}
\def\thetaL{180}
\def\thetaR{0}

% Main circle
\draw[black,line width=1.15pt]
(0,0) circle (\R);

% Subsystem A
\draw[suborange!25,line width=9pt]
(\Astart:\R)
arc[start angle=\Astart,end angle=\Aend,radius=\R];

\draw[suborange,line width=5.2pt]
(\Astart:\R)
arc[start angle=\Astart,end angle=\Aend,radius=\R];

\node[font=\Large,anchor=south]
at (0,2.63) {$A$};

% Antipodal points
\fill (\thetaL:\R) circle (2.2pt);
\fill (\thetaR:\R) circle (2.2pt);

\draw[guidegray,densely dotted,line width=1.1pt]
(\thetaL:\R) -- (\thetaR:\R);

% Blue quasiparticles
\draw[qpblue,line width=2.2pt,-latex]
(\thetaL:\Router)
arc[start angle=180,end angle=162,radius=\Router];

\draw[qpblue,line width=2.2pt,-latex]
(\thetaR:\Router)
arc[start angle=0,end angle=18,radius=\Router];

% Red quasiparticles
\draw[qpred,dashed,line width=2.2pt,-latex]
(\thetaL:\Router)
arc[start angle=180,end angle=198,radius=\Router];

\draw[qpred,dashed,line width=2.2pt,-latex]
(\thetaR:\Router)
arc[start angle=0,end angle=-18,radius=\Router];

% Momentum labels
\node[qpblue,font=\large,anchor=east]
at (-2.95,0.45) {$k$};

\node[qpblue,font=\large,anchor=west]
at (2.95,0.45) {$-k$};

\node[qpred,font=\large,anchor=east]
at (-2.95,-0.45) {$-k$};

\node[qpred,font=\large,anchor=west]
at (2.95,-0.45) {$k$};

\end{tikzpicture}%
}

% ============================================================
% ORIGINAL SLIDE PANEL (e)
% ============================================================

\def\CCpanelE{%
\begin{tikzpicture}[
    line cap=round,
    line join=round,
    scale=0.68,
    every node/.style={transform shape}
]

\path[use as bounding box]
(-3.3,-3.45) rectangle (3.3,3.30);

\def\R{2.50}
\def\Router{2.62}
\def\Astart{55}
\def\Aend{125}

\def\meetTop{90}
\def\meetBottom{270}
\def\dold{18}
\def\dnew{34}

% Main circle
\draw[black,line width=1.15pt]
(0,0) circle (\R);

% Subsystem A
\draw[suborange!25,line width=9pt]
(\Astart:\R)
arc[start angle=\Astart,end angle=\Aend,radius=\R];

\draw[suborange,line width=5.2pt]
(\Astart:\R)
arc[start angle=\Astart,end angle=\Aend,radius=\R];

\node[font=\Large,anchor=south]
at (0,2.63) {$A$};

% Top blue pair
\fill[qpblue,opacity=0.35]
(\meetTop:\R) circle (2.2pt);

% Earlier motion
\draw[qpblue,line width=1.7pt,opacity=0.35,-latex]
(\meetTop:\Router)
arc[start angle=\meetTop,
    end angle={\meetTop+\dold},radius=\Router];

\draw[qpblue,line width=1.7pt,opacity=0.35,-latex]
(\meetTop:\Router)
arc[start angle=\meetTop,
    end angle={\meetTop-\dold},radius=\Router];

% Later motion
\draw[qpblue,line width=2.2pt,-latex]
({\meetTop+\dold}:\Router)
arc[start angle={\meetTop+\dold},
    end angle={\meetTop+\dnew},radius=\Router];

\draw[qpblue,line width=2.2pt,-latex]
({\meetTop-\dold}:\Router)
arc[start angle={\meetTop-\dold},
    end angle={\meetTop-\dnew},radius=\Router];

% Bottom red pair
\fill[qpred,opacity=0.35]
(\meetBottom:\R) circle (2.2pt);

% Earlier motion
\draw[qpred,dashed,line width=1.7pt,opacity=0.35,-latex]
(\meetBottom:\Router)
arc[start angle=\meetBottom,
    end angle={\meetBottom-\dold},radius=\Router];

\draw[qpred,dashed,line width=1.7pt,opacity=0.35,-latex]
(\meetBottom:\Router)
arc[start angle=\meetBottom,
    end angle={\meetBottom+\dold},radius=\Router];

% Later motion
\draw[qpred,dashed,line width=2.2pt,-latex]
({\meetBottom-\dold}:\Router)
arc[start angle={\meetBottom-\dold},
    end angle={\meetBottom-\dnew},radius=\Router];

\draw[qpred,dashed,line width=2.2pt,-latex]
({\meetBottom+\dold}:\Router)
arc[start angle={\meetBottom+\dold},
    end angle={\meetBottom+\dnew},radius=\Router];

% Momentum labels
\node[qpblue,font=\large,anchor=east]
at (-1.55,2.05) {$-k$};

\node[qpblue,font=\large,anchor=west]
at (1.55,2.05) {$k$};

\node[qpred,font=\large,anchor=east]
at (-1.55,-2.05) {$k$};

\node[qpred,font=\large,anchor=west]
at (1.55,-2.05) {$-k$};

\end{tikzpicture}%
}

% ============================================================
% ORIGINAL SLIDE PANEL (f)
% Also used for panel (a)
% ============================================================

\def\CCpanelF{%
\begin{tikzpicture}[
    line cap=round,
    line join=round,
    scale=0.68,
    every node/.style={transform shape}
]

\path[use as bounding box]
(-3.3,-3.45) rectangle (3.3,3.30);

\def\R{2.50}
\def\Router{2.62}
\def\Astart{55}
\def\Aend{125}
\def\thetaL{180}
\def\thetaR{0}

% Main circle
\draw[black,line width=1.15pt]
(0,0) circle (\R);

% Subsystem A
\draw[suborange!25,line width=9pt]
(\Astart:\R)
arc[start angle=\Astart,end angle=\Aend,radius=\R];

\draw[suborange,line width=5.2pt]
(\Astart:\R)
arc[start angle=\Astart,end angle=\Aend,radius=\R];

\node[font=\Large,anchor=south]
at (0,2.63) {$A$};

% Antipodal connection
\draw[guidegray,line width=1.1pt]
(\thetaL:\Router) -- (\thetaR:\Router);

% Left particle
\fill[guidegray] (\thetaL:\R) circle (2.2pt);

\draw[qpblue,line width=2.2pt,-latex]
(\thetaL:\Router)
arc[start angle=180,end angle=198,radius=\Router];

\draw[qpred,dashed,line width=2.2pt,-latex]
(\thetaL:\Router)
arc[start angle=180,end angle=162,radius=\Router];

% Right particle
\fill[guidegray] (\thetaR:\R) circle (2.2pt);

\draw[qpblue,line width=2.2pt,-latex]
(\thetaR:\Router)
arc[start angle=0,end angle=-18,radius=\Router];

\draw[qpred,dashed,line width=2.2pt,-latex]
(\thetaR:\Router)
arc[start angle=0,end angle=18,radius=\Router];

% Momentum labels
\node[qpblue,font=\large,anchor=east]
at (-2.95,-0.55) {$k$};

\node[qpred,font=\large,anchor=east]
at (-2.95,0.55) {$-k$};

\node[qpblue,font=\large,anchor=west]
at (2.95,-0.55) {$-k$};

\node[qpred,font=\large,anchor=west]
at (2.95,0.55) {$k$};

\end{tikzpicture}%
}

% ============================================================
% PANEL WRAPPER
%
% Fixed-width minipages inside a tabular environment.
% This FORCES a 3-column layout.
% ============================================================

\setlength{\fboxsep}{3pt}
\setlength{\fboxrule}{0.45pt}

% #1: subfigure letter, #2: original TikZ drawing, #3: time
\def\CCcell#1#2#3{%
    \fcolorbox{guidegray!45}{white}{%
    \begin{minipage}[t]{0.277\textwidth}%
        \vspace{0.4ex}%
        \noindent{\small\bfseries\color{guidegray}(#1)}\par
        \vspace{1.0ex}%
        \centering
        \resizebox{\linewidth}{!}{#2}\par
        \vspace{0.55ex}%
        {\small\color{black!80}$#3$}\par
        \vspace{0.5ex}%
    \end{minipage}%
    }%
}

% ============================================================
% THREE COLUMNS, TWO ROWS
% ============================================================
\begin{tabular}{@{}c@{\hspace{0.018\textwidth}}c@{\hspace{0.018\textwidth}}c@{}}
    \CCcell{a}{\CCpanelF}{t=0}
    &
    \CCcell{b}{\CCpanelB}{t\simeq t_{\mathrm{drop}}}
    &
    \CCcell{c}{\CCpanelC}{t=\mathcal L/4}
    \\[1.8ex]
    \CCcell{d}{\CCpanelD}{t=\mathcal L/2}
    &
    \CCcell{e}{\CCpanelE}{t=3\mathcal L/4}
    &
    \CCcell{f}{\CCpanelF}{t=\mathcal L}
\end{tabular}

\caption{
Quasiparticle propagation following a quench from the crosscap state.
The orange arc denotes subsystem $A$, while the violet arc, where shown,
denotes its antipodal image $A^M$.
Panels (a)--(f) show successive stages of the dynamics; for unit velocity,
$t_{\mathrm{drop}}=\mathcal L/4-\ell/2$.
}
\label{fig:crossasym:qpp_six_panels}

\end{figure*}

We now turn to the study of the dynamics of the entanglement asymmetry in our crosscap quench. As mentioned above, we cannot directly compute the entanglement asymmetry using the eigenvalues of the $\Gamma_A(t)$ and so we instead look at the behaviour of the R\'enyi 2 entanglement asymmetry as a proxy for this behaviour. We split the discussion into two parts depending on whether the subsystem is smaller or larger than half the total system size. Following this we present the quasiparticle picture prediction for the entanglement asymmetry which works for any replica index allowing us to study the behaviour of $\Delta S[\rho_A(t)]$.

\subsection{Dynamics of the R\'enyi 2 entanglement asymmetry}
\subsubsection{Symmetric initial subsystems}
For $\ell\leq L$ we have seen that the initial R\'enyi entanglement asymmetries in the deformed crosscap states vanish for all $n$. As we will see, this does not remain the case at all times under time evolution. In Figure \ref{fig:crossasym:renyi2_asymmetry_dynamics} we plot the quench dynamics of the $n=2$ entanglement asymmetry obtained through the charged moments by numerically evaluating~\eqref{eq:crossasym:charged_mom_order_n_gaussian} and performing the $\lambda$ integrals. We see that at short time $\Delta S^{(2)}[\rho_A(t)]$ remains zero. At $t=\frac{L-\ell}{4}$  it starts to increase, reaching a maximum at around $t=\mathcal{L}/4$ when it starts to decrease. 
As time progresses $\Delta S^{(2)}[\rho_A(t)]$ experiences oscillations but never drops back to zero. Thus, in contrast to previous studies of the quench dynamics of the entanglement asymmetry, unitary time evolution with a symmetric Hamiltonian causes local symmetry breaking rather than symmetry restoration. Non-decreasing and non-monotonic behaviour of  subsystem asymmetry has previously been seen before in an open system setting and in closed bosonic systems but not in closed fermionic systems~\cite{hammer2026asymmetrydynamicsnonequilibriumsymmetrybreaking,yamashika_superfluid_2d_2025}. 

The times at which the asymmetry starts to increase or decrease coincide with the opposite behaviour in the entanglement entropy indicating a common mechanism underlying their dynamics and therefore an explanation of the observed behaviour. We expand on this in the next section but first consider larger subsystem sizes.

\subsubsection{Asymmetric initial subsystems} 
When $\ell>L$ the reduced density matrix is no longer symmetric and the initial value of the R\'enyi asymmetries are non-zero. We plot the dynamics of the $n=2$ asymmetry in the right panel of Figure \ref{fig:crossasym:renyi2_asymmetry_dynamics} for these sizes of subsystem. We see that indeed $\Delta S^{(n)}[\rho_A]$ starts from a finite value and stays constant for a period of time before increasing and then undergoing periodic revivals. We can note that the asymmetry never drops below its initial value, while the time at which the asymmetry starts to increase is $t=\frac{\ell-L}{4}$ which is the time at which a subsystem of size $2L-\ell$ would start to experience non-trivial entanglement dynamics. Both of these observations, along with the behaviour of the initially symmetric subsystem, can be understood through the quasiparticle picture, which we now describe.

\subsection{Quasiparticle picture}
In free fermion quenches, it has been previously shown that the charged moments admit an interpretation in terms of the quasiparticle picture~\cite{SREE14_qpp_sree_cft_ffs,SREE01Parez_2021,ares2023entanglement}, which can then be used to provide a quasiparticle prediction of the entanglement asymmetry from lowly entangled initial states~\cite{rylands2024microscopic,Chalas_2024,cma-24}. Here we adapt this to study the asymmetry dynamics in quenches starting from crosscap states. To do this we start from the structure of the initial state charged moments given in~\eqref{eq:crossasym:initia_charegd_moment}. The two types of terms present there arise from the pure part of the reduced density matrix which contains $|A_2|/2$ correlated pairs of sites and the mixed term  describing the $|A_1|$ sites whose correlated antipode is not within $A$. Within the context of the quasiparticle picture, $|A_1|$ can be viewed as the initial number of quasiparticles which are shared between $A$ and its complement, i.e. $|A_1|=\ell_{\rm s}(0)$. Likewise, $|A_{2}|/2$ is the number of quasiparticle pairs which are initially completely within the subsystem, as can be seen from Figure~\ref{fig:crossasym:schematic_antipodal}. We denote the number of complete pairs by $\ell_{\rm c}(t)$. Since the average number of quasiparticles does not change throughout the quench and their density is homogeneous we have the sum rule $\ell=\ell_{\rm s}(t)+2\ell_{\rm c}(t)$ and hence
\begin{eqnarray}
    \ell_{\rm c}(t)=\frac{\ell-\ell_{\rm s}(t)}{2}~.
\end{eqnarray}
This leads us to the following quasiparticle prediction for the time evolution of the  charged moments
\begin{eqnarray}\label{eq:crossasym:time_evolved_charge_mom}
      Z_n(\vec{\lambda},t)=\prod_{j=1}[1-\theta_ \mu +e^{2i(\lambda_j-\lambda_{j+1})}\theta_ \mu]^{\frac{\ell-\ell_{\rm s}(t)}{2}} \big[(1-\theta_ \mu)^n+\theta_ \mu^n\big]^{{\ell_{\rm s}(t)}}~.
\end{eqnarray}
Upon setting $\lambda_j=0$ $\forall j$, the term describing complete pairs disappears and we can recover the quasiparticle prediction for the R\'enyi entanglement entropies~\eqref{eq:crossasym:ent_ent_qpp}. The above expression is simple enough  that the $\lambda$ integrals can be performed exactly and we arrive at the following expression for the R\'enyi entanglement asymmetries.
Using~\eqref{eq:crossasym:initia_charegd_moment} in~\eqref{eq:crossasym:chraged_moment_asymmetry} and performing the same changes of variables~\cite{rylands2024microscopic} we find that
\begin{eqnarray}\label{eq:crossasym:time_evolved_renyi_asymm}
    \Delta S^{(n)}[\rho_A(t)]=\frac{1}{1-n}\log\left[\sum_{q}J(q,t)^n\right]
\end{eqnarray}
and in the replica limit
\begin{eqnarray}\label{eq:crossasym:time_evolved_asymm}
    \Delta S[\rho_A(t)]=-\sum_{q} J(q,t)\log J(q,t)~,
\end{eqnarray}
where now
\begin{eqnarray}
     J(q,t)&=&\int_{-\pi}^\pi\frac{{\rm d}\lambda}{2\pi}\,\big(1-\theta_ \mu+e^{2i\lambda}\theta_ \mu\big)^{\frac{\ell-\ell_{\rm s}(t)}{2}}e^{-i\lambda q }.
\end{eqnarray}
In Figure~\ref{fig:crossasym:renyi2_asymmetry_dynamics} we compare this quasiparticle prediction for $n=2$ with the exact numerics finding perfect agreement. 
Here we have assumed that $(\ell-\ell_{\rm s}(t))/2$ is an integer. Since the quasiparticle picture generally yields a noninteger effective number of complete pairs, $\ell_{\rm c}(t)=(\ell-\ell_{\rm s}(t))/2$, we introduce the following discretization:
\begin{equation}
  J(q,t)  =\begin{cases}      \binom{\lfloor{\frac{\ell-\ell_{\rm s}(t)}{2}\rfloor}}{q/2}\theta_ \mu^{q/2}(1-\theta_ \mu)^{\lfloor\frac{\ell-\ell_{\rm s}(t)}{2}\rfloor-\frac{q}{2}}~&~\text{for $q$ even}\\
         0~&~\text{for $q$ odd}
     \end{cases}~.
\end{equation}
This prescription provides an effective approximation to the quasiparticle counting, preserving the normalization and positivity of the resulting probability distribution. Since the floor function changes the effective number of complete pairs by less than one, it does not affect the leading logarithmic behaviour of the asymmetry for large $\ell_{\rm c}(t)$. We have also verified numerically that this approximation reproduces the leading behaviour of the exact results in the large-subsystem limit.

By construction this reduces to the correct initial state expression at $t=0$. The synchronisation of the entanglement asymmetry and the entanglement entropy witnessed in the exact numerics of $\Delta S^{(2)}[\rho_A(t)]$ is now readily explained. Each quantity is governed by the number of quasiparticle pairs which are shared between $A$ and its complement. A shared pair increases the entanglement but decreases asymmetry while an unshared pair does the opposite. The same connection is present in lowly entangled quenches to integrable models where relaxation is driven by the spreading of entanglement and concomitantly the restoration of the subsystem symmetry. Here, the initially decreasing entanglement drives the generation of symmetry breaking within the subsystem. 
This behaviour is illustrated in Figure~\ref{fig:crossasym:qpp_six_panels}. In panel (a) the initial state is depicted and the pairs are antipodally shared meaning that entanglement is large, but the state is locally symmetric since there are no pairs within $A$.
The pairs start to move towards each other, where we see that we have blue and red colours for the pairs which emanate from the antipodal points.
At time $t=\frac{\mathcal{L}}{4}$, keeping track of particles with velocity $v=1$, an entangled pair meets at the same spatial point, which is the lowest amount of entanglement we can have, corresponding to the maximum of the asymmetry, since all pairs inside $A$ are complete.
In this sense we see the compatibility between the two quasiparticle pictures, and that which is present in quenches from lowly entangled states~\cite{ares2023entanglement,rylands2024microscopic}.

For a large enough number of complete pairs ($\ell_c\gg1$) the asymmetry can be well approximated along the lines of \eqref{eq:crossasym:approx_asymmetry} as
\begin{eqnarray}\label{eq:approx_asymm}
    \Delta S^{(n)}[\rho_A(t)]\simeq \frac{1}{2}\log[ 2\pi n^{\frac{1}{n-1}} \ell_{\rm c}(t)\theta_ \mu(1-\theta_ \mu)]~\\
    \Delta S[\rho_A(t)]\simeq \frac{1}{2}\log[2 \pi e \ell_{\rm c}(t)\theta_ \mu(1-\theta_ \mu)]~
\end{eqnarray}
which is valid for $\ell>L$ and for $t>(L-\ell)/4$ when $\ell<L $. 

% From the quasiparticle picture, we can see that we have an emergent effective length which appears through $\ell_c(t)$ that counts the complete pairs in $A$.
% Moreover, in the case of $\ell>L$ we can observe from the numerics that the asymmetry does not drop below its initial value, which is viewed by the fact that $\ell_c(t)$ has the same behaviour.
% Thus, since the subsystem is large enough, it will always break the symmetry at or above its initial value.

\subsection{Long time average}
\begin{figure}
    \centering

    \includegraphics[width=0.45\linewidth]{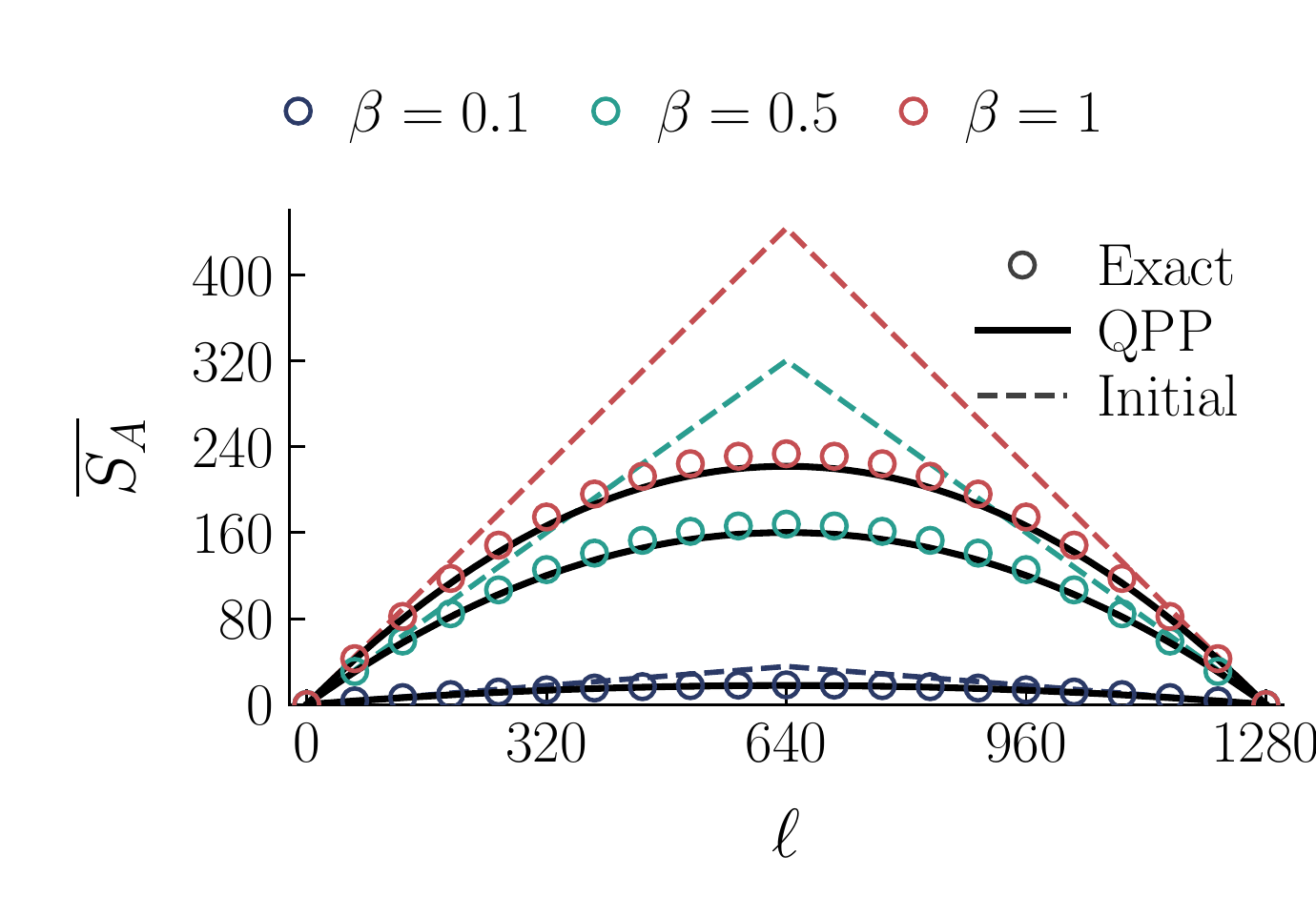}
     \includegraphics[width=0.45\linewidth]{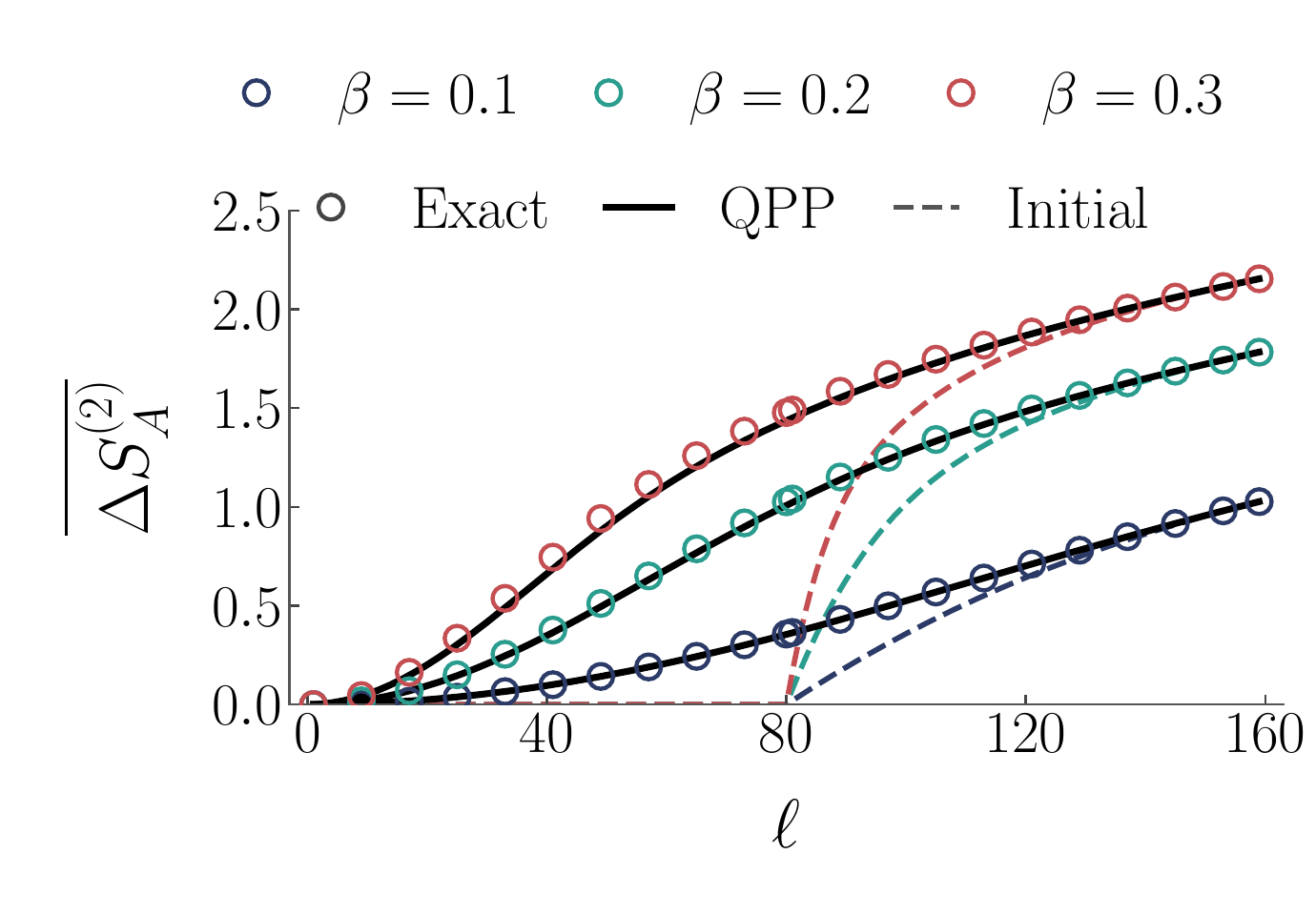}
    \caption{Left: The long time averaged value of $S[\rho_A]$ for $\beta=0.1,0.5,1$ (solid lines). For comparison plot the initial state values (dashed lines). Right: The long time averaged R\'enyi entanglement asymmetry $\overline{\Delta S^{(2)}[\rho_A]}$ for the same values of $\beta=0.1,0.2,0.3$ (solid) with the initial values for comparison.  Left panel considers a chain of size $\mathcal{L}=1280$, while the second one considers a chain of $\mathcal{L}=160$, the time average is taken in the window $t\in[5\mathcal{L},20\mathcal{L}]$.}
    \label{fig:crossasym:time_avg_entropy_and_asymmetry}
\end{figure}
In order to witness non-trivial dynamics we must study the system on times scales which scale like $L$. Therefore, as a result of finite size revivals, the system is not expected to show any relaxation toward a local equilibrium state as would be the case when considering a quench from a lowly entangled state. We can instead explore the long time average of the subsystem properties. The average over an observable $\mathcal{O}(t)$ is defined as 
\begin{eqnarray}
    \overline {\mathcal{O}}=\lim_{T\to \infty}\frac{1}{T}\int^{T}_0 \!\!{\rm d} t \,\mathcal{O}(t)~.
\end{eqnarray}
Using the result of~\eqref{eq:crossasym:ent_ent_qpp} we obtain the long time averaged entanglement 
\begin{eqnarray}
     \overline {S^{(n)}[\rho_A]}&=&\lim_{T\to \infty}\frac{s^{(n)}_ \mu}{T}\int^{T}_0 \!\!{\rm d} t \,\ell_{\rm s}(t)=s^{(n)}_ \mu \frac{\ell(2L-\ell)}{2L}~,
\end{eqnarray}
which is depicted in solid lines in the left panel of Figure~\ref{fig:crossasym:time_avg_entropy_and_asymmetry}, where we see the excellent agreement with the exact time averaged values depicted with open circles.
We see that $\overline {S^{(n)}[\rho_A]}$ takes the same form as the previously derived expression for the undeformed crosscap states~\cite{chalas2024quenchdynamicsentanglementcrosscap}. It can be interpreted by recalling that $s^{(n)}_ \mu$ is the contribution to the entanglement of a correlated pair of quasiparticles which are shared between $A$ and $\bar{A}$ while the other factor is merely the average number of shared pairs  if all quasiparticles are uniformly distributed over the system. 

Likewise, using the quasiparticle prediction \eqref{eq:approx_asymm} for the asymmetry, we have the time averaged expressions
\begin{eqnarray}
    \overline{\Delta S^{(n)}[\rho_A]}&=&\frac{1}{2}\log [2\pi n^{\frac{1}{n-1}} \theta_ \mu (1-\theta_ \mu)]+\lim_{T\to \infty}\frac{1}{2T}\int^{T+\tau}_\tau \!\!{\rm d} t \,\log l_{\rm c}(t) \\
    \overline{\Delta S[\rho_A]}&=&    \frac{1}{2}\log [2\pi e \theta_ \mu (1-\theta_ \mu)]+\lim_{T\to \infty}\frac{1}{2T}\int^{T+\tau}_\tau \!\!{\rm d} t \,\log l_{\rm c}(t)~,
    \label{eq:crossasym:time_avgd_asymmetries}
\end{eqnarray}
where we have introduced a short time cutoff $t_{\rm drop}<\tau\ll T$ which allows us to use the approximate expressions~\eqref{eq:approx_asymm}. Using these we see a very good agreement between the quasiparticle picture average and the numerical one in Figure~\ref{fig:crossasym:time_avg_entropy_and_asymmetry}, which also depicts the initial state value shown with the dashed lines.
This result shows the symmetry breaking that results from the symmetry preserving evolution of an initially locally symmetric state.
We see that for finite subsystems we have a non-zero value for the asymmetry showing the presence of  a non-zero long-time-averaged symmetry breaking, which is not present at the start of the quench.

Interestingly, for large subsystem and system sizes, there seems to be an approximate equality between the quenched and annealed averages.
In particular, let us denote by $\widehat{\rho_A}$ the reduced density matrix constructed so that it contains the correct long time average number of shared and complete quasiparticle pairs~ \cite{rottoli2024entanglement,Travaglino_2025}.  We find that $ S^{(n)}[\widehat{\rho_A}]=\overline{S^{(n)}[\rho_A]}$, meaning that the quenched versus annealed averages are approximately equal in the case of the entanglement entropy, but for the asymmetries, it is not obvious that the equations~\eqref{eq:crossasym:time_avgd_asymmetries} are equal to 
\begin{eqnarray}
       \Delta S^{(n)}[\widehat{\rho_A}]&=&\frac{1}{2}\log \Big[\pi n^{\frac{1}{n-1}}\frac{\ell^2}{2L}\theta_ \mu(1-\theta_ \mu)\Big] \\\quad \Delta S[\widehat{\rho_A}]&=&\frac{1}{2}\log \Big[\pi e\frac{\ell^2}{2L}\theta_ \mu(1-\theta_ \mu)\Big]~.
       \label{eq:crossasy_state_avgd_asymmetries}
\end{eqnarray}
We have numerically verified that for large subsystem and system sizes  equations~\eqref{eq:crossasy_state_avgd_asymmetries}  and~\eqref{eq:crossasym:time_avgd_asymmetries} agree at leading logarithmic order. In other words, this suggests that the time average of the entanglement asymmetry and the asymmetry of the time-averaged density matrix exhibit the same leading scaling behaviour, with differences appearing only at subleading orders. Moreover, this hints that the leading contribution to the asymmetry is determined by the average number of complete quasiparticle pairs within the subsystem. Interestingly, this is reminiscent of the behaviour of the entanglement entropy, whose leading contribution is governed by the average number of quasiparticle pairs shared between $A$ and $\bar{A}$.

% This means that alternatively, we could instead take the average on the reduced density matrix at the level of the quasiparticle picture by considering it being made up of the time averaged number of shared and complete pairs. For the entanglement, exchanging of these annealed and quenched averages does not change the final result, however this is not immediately understood if it is the case for the entanglement asymmetry. Denoting the time averaged state by  $\overline{\rho_A}$. we have $ S^{(n)}[\overline{\rho_A}]=\overline{S^{(n)}[\rho_A]}$ but
% \begin{eqnarray}
%        \Delta S^{(n)}[\overline{\rho_A}]&=&\frac{1}{2}\log \Big[\pi n^{\frac{1}{n-1}}\frac{\ell^2}{\mathcal{L}}\theta_ \mu(1-\theta_ \mu)\Big] \\\quad \Delta S[\overline{\rho_A}]&=&\frac{1}{2}\log \Big[\pi e\frac{\ell^2}{\mathcal{L}}\theta_ \mu(1-\theta_ \mu)\Big]~,
% \end{eqnarray}
% where we see that we have taken the average of the number of complete pairs within $A$ which is $\frac{\ell^2}{\mathcal{L}}$.
% We plot the long time averaged quantities as a function of system size in Figure \ref{fig:crossasym:time_avg_entropy_and_asymmetry}. 

\section{Conclusions}
\label{sec:crossasym:conclusions}
In this paper we have introduced the class of fermionic \textit{charge deformed crosscap states}.
We have studied the interplay between entanglement, symmetry and symmetry breaking, with respect to the $U(1)$ fermionic number charge in and out of equilibrium.
These are Gaussian, squeezed states featuring antipodal correlations the magnitude of which is deformed by the fermion number operator.
The charge deformation allows these states to have controllable entanglement and symmetry breaking, as measured by the entanglement asymmetry.

Charge deformed crosscap states exhibit particularly interesting features of local symmetry breaking, which unveil how fine-tuned long-range entanglement can create complex symmetry breaking patterns.
As is known, the antipodal correlations endow the state with a static volume law for the entanglement entropy with its value tuned by the magnitude of the deformation.
This entanglement structure gives these states  intricate symmetry breaking properties.
As is manifest from their definition, the charge deformed crosscap states globally break particle number symmetry since they have symmetry breaking excitations which are antipodally correlated.
Our analysis shows that the charge deformed crosscap states, for subsystem sizes smaller than half of the total system size $\ell<\mathcal{L}/2$, the subsystems are locally symmetric.
For subsystems $\ell>\mathcal{L}/2$ the subsystem starts to break the symmetry in an amount controlled by the charge deformation.
This signals that the information that the global state breaks the symmetry is accessed for subsystem sizes $\ell\geq L$.
The symmetry breaking pattern of the crosscap state is directly caused by its antipodal entanglement pattern.

The case of the maximally entangled crosscap state gives a  volume law which looks similar to the Page curve of entanglement which can be seen in random states~\cite{page1993average}, albeit with the caveat that in that case one has to take an average over a state ensemble and the result holds true only in the thermodynamic limit.

Locally, a subsystem of the crosscap state in the case of $\ell<\mathcal{L}/2$, is in a symmetric product Gibbs state  and thus does not contain any symmetry breaking pure part locally.
As soon as we include two entangled sites within the subsystem, entering the regime of $\ell\geq\mathcal{L}/2$ , the subsystem acquires moreover a pure part, which is symmetry breaking.
With a growing subsystem, increasingly more entangled pairs are included and symmetry is increasingly broken locally.
This terminates when we consider the full system which gives the maximal symmetry breaking for that state.
The increase here at $\ell=\frac{\mathcal{L}}{2}$ does not show signs of a discontinuity at the limit of $\ell,\mathcal{L}\rightarrow\infty$.

This behaviour of symmetry breaking at $\ell\geq\frac{\mathcal{L}}{2}$ is reminiscent of the symmetry breaking study of Haar random states studied  in~\cite{ampc-23,Russotto_2025}.
These states locally respect the symmetry, in the thermodynamic limit, for $\ell<\mathcal{L}/2$, due to the Hayden-Preskill inequality~\cite{hayden_preskill_ineq_2007}, while there is a sharp discontinuity at $\ell=\mathcal{L}/2$ after which the subsystem is symmetry breaking.
Again, we should restate the caveat that the result for Haar random states holds for an ensemble of states in the thermodynamic limit.
What we see here is that the entanglement behaviour of such states is reminiscent of the Page result~\cite{page1993average}, but the asymmetry calculation sets them clearly apart~\cite{ampc-23,Russotto_2025}, even in the thermodynamic limit.
Although they share some common traits, they have important differences, which are expected since the crosscap states are finely tuned entangled antipodal pair states~\cite{yoneta_eap_PhysRevLett.133.170404}.

The dynamical evolution is particularly interesting, as it shows that a symmetry preserving subsystem can be time evolved to be symmetry breaking.
This is quite unexpected, since in a standard  quench the state starts from symmetry breaking and subsequently the symmetry is restored~\cite{ares2023entanglement}.
This behaviour was captured in excellent agreement with exact numerics by the antipodal quasiparticle picture of the crosscap states~\cite{chalas2024quenchdynamicsentanglementcrosscap} which was adapted for the entanglement asymmetry.
In this case, we see that the quasiparticle picture can explain a quite counter-intuitive behaviour of symmetry breaking in a symmetric evolution starting from a symmetric subsystem.

It is worth emphasizing that the local symmetry breaking observed in our quench does not correspond to the generation of additional symmetry breaking at the global level. Indeed, since the dynamics preserves the total charge, the global entanglement asymmetry remains constant throughout the evolution. Rather, the dynamics redistributes the initially nonlocal symmetry-breaking coherence, encoded in correlations between antipodal sites, making it progressively accessible to local subsystems. From this perspective, our results illustrate how the spatial structure of entanglement determines not only the amount of locally accessible quantum information, but also whether the symmetry-breaking properties of a global quantum state can be detected locally.

A natural question for future investigations is how robust this unconventional symmetry-breaking dynamics is against perturbations of the initial state and the subsequent evolution. In particular, it would be interesting to determine whether the redistribution of local entanglement asymmetry found here persists when antipodal correlations are weakened or the free-fermion dynamics is replaced by charge-conserving interacting integrable or chaotic dynamics. For evolutions that do not conserve the charge, one could also ask whether local asymmetry can be generated dynamically. This would allow us to establish whether the observed behaviour is a special consequence of the fine-tuned crosscap structure or represents a more general mechanism through which unitary, symmetry-preserving dynamics can reveal initially hidden symmetry breaking. More broadly, it would be interesting to identify which properties of the initial entanglement structure determine whether local symmetry restoration or symmetry breaking takes place during nonequilibrium evolution.

There is another direction we find appealing: generalizing the construction to non-Abelian \cite{Russotto_2025} and non-invertible symmetries \cite{fossati_2025_noninvertible_asymm}.
Indeed, our mechanism fundamentally involves the way symmetry-breaking information is distributed among entangled pairs. For more general symmetries, this information may have a considerably richer structure, potentially leading to qualitatively different forms of local symmetry breaking.

Finally, it would be very interesting to understand the symmetry breaking dynamics of the entangled multipodal states introduced in~\cite{chalas2025entanglementevolutionentangledmultipodal}.
In this case the entanglement patterns are different from the crosscap state. In particular one can have an increase of the entropy in time, not available in the crosscap states.
Symmetry breaking in these settings could lead to very intriguing behaviours.
Moreover, this shows the possibility of engineering interesting behaviours of symmetry breaking in and out of equilibrium, by tuning the entanglement structure of the initial states in a quench.
Moving in this direction one could define even more complicated initial states, and subsequently understand how to fine-tune their entanglement and symmetry breaking dynamics to generate interesting physics.

\paragraph{Acknowledgements ---}
We thank Filiberto Ares for interesting discussions. KC and PC  acknowledge support by the ERC-AdG grant MOSE No. 101199196.

\bibliographystyle{ytphys}
\bibliography{mybib-thesis}
\end{document}